\documentclass[aps,prd,preprint,
eqsecnum,nofootinbib, 10pt]{revtex4-2}
\usepackage{graphicx,xcolor} 
\usepackage{amsmath,amssymb}
\usepackage{braket,slashed}

\allowdisplaybreaks

\usepackage[pdfstartview=XYZ,
bookmarks=true,
colorlinks=true,
linkcolor=blue,
urlcolor=blue,
citecolor=blue,
bookmarks=true,
linktocpage=true, 
hyperindex=true]{hyperref}
\usepackage{orcidlink}
\usepackage[normalem]{ulem}

\newcommand{\half}{{\frac{1}{2}}}
\newcommand{\La}{\mathcal{L}}

\newcommand{\Ha}{\mathcal{H}}

\newcommand{\Real}{\operatorname{Re}}
\newcommand{\Imag}{\operatorname{Im}}

\newcommand{\Le}{\mathbb{L}}
\newcommand{\Ri}{\mathbb{R}}
\newcommand{\hc}{\text{h.c.}}

\begin{document}
\title{Torsional four-fermion interaction for Majorana neutrinos}
\author{Indrajit Ghose \orcidlink{0000-0002-8561-4954}}
\email{ghose.meghnad@gmail.com}
\author{Amitabha Lahiri \orcidlink{0000-0001-8113-6345}}
\email{amitabha@bose.res.in}
\affiliation{S. N. Bose National Centre for Basic Sciences\\
	Block JD, Sector 3, Salt Lake 700106, India.}

\begin{abstract}

Fermions generate spacetime torsion, which can be eliminated, leaving behind an effective four-fermion interaction. This term will contribute an effective mass for neutrinos propagating through matter, similar to the Wolfenstein term coming from electroweak interactions. 
When the neutrinos are Majorana fermions and become massive via the Type I SeeSaw mechanism, there can be additional effects due to sterile neutrinos interacting with all fermions via the torsion-induced term, as well as due to the presence of new mixing parameters. We consider different scenarios with one sterile and one or two active neutrinos -- when the torsional interaction is diagonal in the mass basis and when it is not -- and analytically find these modifications. The T violation in the 2+1 scenario is discussed for some specific mixing between the torsion fields and the mass fields.
\end{abstract}

\maketitle
\newpage

\section{Introduction}

{Neutrinos are massless in the Standard Model. Experiments have shown however that neutrinos undergo flavor oscillations, which is not a prediction of the Standard Model but is considered to be strong evidence that neutrinos have small but non-zero masses, not all equal.} 
There {are several models which} naturally produce very small neutrino masses, for example the SeeSaw type models~\cite{Minkowski:1977sc, Yanagida:1980xy, Mohapatra:1979ia, Schechter:1980gr, Foot:1988aq, King:2003jb, deGouvea:2016qpx} or Radiative mass models~\cite{Cai:2017jrq, Babu:2019mfe} that produce masses {via loop corrections} which are naturally small. GUTs and supersymmetric theories naturally incorporate some combination of different SeeSaw mechanisms as a simple explanation of the smallness of neutrino masses \cite{Mohapatra:1987fs, Coriano:2003ui, Masiero:2005ua, Senjanovic:2005sf, Bajc:2006ia, Buchmuller:2007zd, Ellis:2007wz, Senjanovic:2009kja, Khalil:2015naa, Bjorkeroth:2015tsa, FileviezPerez:2018dyf, Antusch:2022afk}, which makes it a promising candidate to look for in particle colliders \cite{Arhrib:2009mz, Pattanaik:2026mmq, Zhou:2026hdt, Delepine:2026tbh, Batell:2026avi, Cvetic:2026pyt}. Any mechanism that produces masses for neutrinos is of course beyond the Standard Model (BSM). In this paper we will look at how a completely different kind of BSM physics, namely, spacetime torsion and its interaction with fermions, affects the SeeSaw mechanism of neutrino masses.

Specifically, we will consider the Type I SeeSaw mechanism. The SeeSaw mechanism writes the Lagrangian in terms of two types of fields and their charge conjugates. One type is that of left-chiral fields of active neutrinos which form SU(2) doublets with the charged leptons and thus interact with the gauge bosons. The other type consists of right-chiral fields of sterile neutrinos which are gauge singlets. The neutrino mass spectrum comprises a few light neutrinos and $N$ heavy neutrinos, where $N$ is the number of sterile neutrinos included in the theory. The sterile neutrinos are singlets under all gauge groups, but of course they still feel the spacetime geometry in the form of gravitation and torsion.

A natural framework to couple fermions to spacetime geometry is the Einstein-Cartan-Sciama-Kibble (ECSK) theory~\cite{Cartan:1923zea, Cartan:1924yea, Kibble:1961ba, Sciama:1964wt, Hehl:1974cn, Hehl:1976kj, Soffel:1979ma, Bleecker:1990jy,  Hehl:2007bn,  Poplawski:2009fb, Gasperini:2013cru, Mielke:2017nwt, Chakrabarty:2018ybk}. In ECSK theory, which is basically ISO(3,1) gauge theory, the spacetime has curvature as well as torsion, sourced by the fermions themselves. In minimal ECSK, the fermion current that couples to the torsion field is an axial current. However, it was realized that since left- and right-chiral fermions are in distinct representations of SO(3,1), the torsion field can couple to them independently~\cite{Chakrabarty:2018ybk, Barick:2023qjq}. In a series of works, this non-universal chiral coupling is explored in much more detail, including event-level analysis of the sizes of such couplings~\cite{Chakraborty:2024zek, Choudhury:2024vzx, Ghose:2025tgc, Barick:2025iwn}. In those papers, the effects of the Spin-Torsion interaction were considered without considering the origin of the mass. This raises the question what happens if a specific mass mechanism is considered, which we will investigate in this work for the type I SeeSaw mechanism. {The SeeSaw mechanism predicts that the neutrinos are Majorana-type fermions. We will investigate what happens to a Majorana fermion in the presence of the torsional interaction. We also consider the possibility that the coupling between torsion and the fermions is not diagonal in the mass basis. As experiments like DUNE plans to measure CP phases, we explore the contribution of Majorana neutrinos in the CP or T violation. The goal of this work will be to understand the effects of spacetime torsion in the aforementioned setting on the effective dispersion relation in matter and CP violation.}


This paper is arranged as follows. We start by recapitulating the SeeSaw mechanism of type I in Sec.~\ref{sec:seesaw} and derive the approximate flavor-mass mixing matrix. In Sec.~\ref{sec:ecsk}, we briefly talk about the ECSK formalism in the presence of the chiral torsion. In Sec.~\ref{sec:qft}, we discuss the Hamiltonian of the flavor evolution of the neutrinos, including both helicities. In the ultrarelativistic limit, the two helicities decouple. We derive the free Hamiltonian, and the matter potentials in the Second Quantization picture. In Sec.~\ref{sec:1p1}, we talk about the effective mass squared difference in matter in the scenario when only one active neutrino is present with one sterile neutrino. In  
Sec.~\ref{sec:2p1}, we discuss the change in the effective mass squared difference in the presence of matter for the 2+1 model, i.e., when two active neutrinos are present with only one sterile neutrino. We then conclude with a summary and discussion of open problems.

\section{Type I SeeSaw mechanism}
\label{sec:seesaw}

Let us quickly go over the fundamentals of the Type I SeeSaw mechanism -- more details can be found in~\cite{Minkowski:1977sc, Yanagida:1980xy, Mohapatra:1979ia, Mohapatra:1998rq, Giunti:2003dg}. The Standard Model contains only the left-chiral fields of the active neutrinos, which will be denoted $\nu_{\alpha}\,,$ with $\alpha \in \{e, \mu, \tau\}\,$. In the SeeSaw mechanism, we introduce a set of $N$ right-chiral fermion fields which are gauge singlets and thus do not appear in the weak currents. These are known as sterile neutrinos and will be denoted by $N_{\beta}\,,$ with $ \beta \in \{1, 2, \cdots N\}\,. $ 

Let us consider the possible mass terms for the neutrinos. These will contain the active neutrino fields, which are always left-chiral, and the sterile neutrino fields, which are right-chiral. Dirac type mass terms for the neutrinos arise from coupling with the Higgs field and contain $\bar{\nu}N$ and its Hermitian conjugates, while  Majorana type mass terms are of the form $\bar{N}^{c}N$\,, where the superscript $c$ indicates the charge conjugate of the field. 
Then the most general mass term for these neutrino fields can be written as
\begin{align}
    \La_{\operatorname{mass}} =& -\frac{1}{2}\begin{pmatrix}\bar{\nu} & \bar{N}^c\end{pmatrix}\begin{pmatrix}0 & M_{D}\\ M^{T}_{D} & M_{M}\end{pmatrix}\begin{pmatrix}\nu^c \\ N\end{pmatrix} + \hc \notag \\
    =& -\frac{1}{2}\bar{\Psi}_{\Le}\begin{pmatrix}0 & M_{D}\\ M_{D}^{T} & M_{M}\end{pmatrix}\Psi^{c}_{\Le} + \hc,
\end{align}
where we have written 
$\Psi_{\Le} = \begin{pmatrix}\nu_{\Le} \\ N^{c}_{\Ri}\end{pmatrix}$\,.  Because of charge conjugation, $N^{c} :=(N_{\Ri})^{c}$ is a left-chiral field, so $\Psi$ is a multiplet of left-chiral flavor fields. 
The matrix $M_{D}$ contains the Dirac type 
masses and the matrix $M_{M}$ contains all the Majorana type masses.
%
Since the elements of $M_{D}$ are in general complex and we can always choose the $M_{M}$ with all real entries, 
the mass matrix is complex symmetric.  Any complex symmetric matrix can be reduced to a real diagonal matrix through the Takagi-Autonne (TA) decomposition~\cite{Houde:2024mkj, takagi1924algebraic}. The entries of the diagonal matrix will give the masses of the neutrinos. Through TA decomposition we will have
\begin{align}
    \tilde{U}\begin{pmatrix}\mu & 0\\ 0 & M\end{pmatrix}\tilde{U}^{T} = \begin{pmatrix}0 & M_{D}\\ M_{D}^{T} & M_{M}\end{pmatrix}, \label{eq:diagonal}
\end{align}
where $\mu$ is a diagonal matrix containing the masses of the lighter neutrinos and $M$ is also a diagonal matrix, but with very large masses on the diagonal. We will further work in the SeeSaw limit~\cite{deGouvea:2006gz} which is defined as $|M_{D}| \ll M_{M}$. The inequality is to be understood as saying that magnitudes of the non-vanishing  entries of $M_{D}$ are very small compared to the non-vanishing entries of the $M_{M}$. In order to find the diagonalizing matrix of the mass Lagrangian under this approximation, we now follow~\cite{deGouvea:2006gz, Giunti:2007ry} and write the ansatz 
\begin{align}
    \tilde{U} = \begin{pmatrix}1 - \frac{1}{2}\Theta \Theta^{\dagger} & \Theta\\ -\Theta^{\dagger} & 1 -\frac{1}{2}\Theta^{\dagger} \Theta \end{pmatrix}\begin{pmatrix}U & 0\\ 0 & 1\end{pmatrix} = \begin{pmatrix}(1 - \frac{1}{2}\Theta \Theta^{\dagger})U & \Theta \\ -\Theta^{\dagger}U & 1 - \frac{1}{2}\Theta^{\dagger} \Theta\end{pmatrix}\,, \label{eq:ansatz}
\end{align}
where $U$ is a unitary matrix, while $\tilde{U}$ is unitary up to $\mathcal{O}(\Theta^{3})$ as can be seen by direct calculation. Then from Eqs.~\eqref{eq:diagonal} and \eqref{eq:ansatz} and keeping terms up to $\mathcal{O}(\Theta^{2})$, we have
\begin{align}
    U \mu U^{T} = - \Theta M \Theta^{T}, \qquad \Theta M = M_{D}, \qquad M_{M}\left(1-\frac{1}{2}\Theta^{\dagger}\Theta\right) = M\,.
\end{align}
This leads to the well-known relation
\begin{align}
    U \mu U^{T} = - M_{D} M^{-1} M_{D}^{T}\,,\label{diagonal-mu}
\end{align}
producing light masses from the SeeSaw mechanism. {Eq.~\eqref{diagonal-mu} shows the lighter neutrinos are light because the sterile neutrinos are heavy.} 
Writing the mass Lagrangian in terms of the full Majorana field $\Psi_{m} = \Psi_{m\Le} + \Psi^{c}_{m\Le}$,  we have
\begin{align}
    \La_{\operatorname{mass}} =& -\frac{1}{2}\bar{\Psi}_{\Le}\tilde{U}\begin{pmatrix}\mu & 0\\ 0 & M\end{pmatrix}\tilde{U}^{T}\Psi^{c}_{\Le} + \hc = -\frac{1}{2}\bar{\Psi}_{m\Le}\begin{pmatrix}\mu & 0\\ 0 & M\end{pmatrix}\Psi^{c}_{m\Le} + \hc \nonumber \\
    =& -\frac{1}{2}\bar{\Psi}_{m}\begin{pmatrix}\mu & 0\\ 0 & M\end{pmatrix}\Psi_{m} = -\frac{1}{2}\bar{\Psi}_{m}M_{\text{diag}}\Psi_{m}\,
    \label{eq:mass_lagrangian}
.\end{align}
Here $\Psi_{m} = \Psi_{m\Le} + \Psi^{c}_{m\Le}$ is a Majorana field, $\Psi^{c}_{m} = \Psi_{m}$. Hence, neutrinos which gain mass through the Type I SeeSaw mechanism are Majorana fermions in nature. From the mass matrices $M_{D}$ and $M_{M}$, we can calculate the flavor-mass mixing matrix $\tilde{U}$. 

\section{The spin-torsion interaction from the ECSK formalism}
\label{sec:ecsk}
{As we mentioned earlier, the sterile neutrino interacts only with the spacetime --- with gravitation via its energy and with torsion via its spin.  The main purpose of this paper is to look at how the SeeSaw mechanism is affected by torsion generated by the fermions themselves. We start by briefly discussing how torsion arises from fermions in curved spacetime. For more details, we refer the reader to~\cite{Chakrabarty:2019cau, Ghose:2023ttq}.}

In the Einstein-Cartan-Sciama-Kibble (ECSK) formalism~\cite{Cartan:1923zea, Cartan:1924yea, Kibble:1961ba, Sciama:1964wt, Hehl:1974cn, Hehl:1976kj,  Hehl:2007bn,  Poplawski:2009fb, Gasperini:2013cru, Mielke:2017nwt, Chakrabarty:2018ybk},  the $\gamma$ matrices are defined on an ``internal'' flat space by $\left[\gamma^a, \gamma^b\right]_{+} = 2\eta^{ab}$\,. The curved spacetime is related to the internal flat space through the ``tetrad" fields $e^\mu_a$ and the inverse tetrad or ``co-tetrad" fields $e^a_\mu$\,, 
\begin{align}
	\eta_{{ab}}e^a_\mu e^b_\nu = g_{\mu\nu}\,, \quad g_{\mu\nu}e^\mu_a e^\nu_b = \eta_{ab}\,,
	 \quad e^\mu_a e^a_\nu = \delta^\mu_\nu\,.
\end{align}
Here $\mu, \nu, \lambda, \cdots$ denote spacetime indices and $a, b, c, \cdots$ denote internal indices. Spacetime indices are lowered and raised with $g$\,, while internal indices are raised and lowered with $\eta$\,. 

The connection has two types of components, $\Gamma^\lambda{}_{\mu \nu}$ for the spacetime and the spin connection $	A_{\mu}{}^{a}{}_{b}$ for the internal space. 
Furthermore, the connection is assumed to be tetrad-compatible, i.e., 
\begin{align}
\nabla_\mu e^a_\nu = 0 \quad \Rightarrow \quad e^\lambda_a \partial_\mu e^a_\nu + 	A_{\mu}{}^{a}{}_{b} e^b_\nu e^\lambda_a - \Gamma^\lambda{}_{\mu \nu} = 0\,.
	\label{tetrad-postulate}
\end{align}
This relation is sometimes referred to as the \textit{tetrad postulate}. The connection components $\Gamma^{\lambda}{}_{\mu \nu}$ are not \textit{a priori} assumed to be symmetric --- the connection is not torsion-free in the presence of fermions. The spinor derivative under minimal coupling scheme is
\begin{equation}\label{Dirac-operator}
	D_\mu\psi = \partial_\mu\psi -\frac{i}{4} A_\mu{}^{ab} \sigma_{ab}\psi\,, \qquad \sigma_{ab} = \frac{i}{2}\left[\gamma_a\,, \gamma_b\right]_{-}\,.
\end{equation}
Let us therefore split the spin connection as 
\begin{equation}\label{split}
	A_{\mu}{}^{ab} = \omega_{\mu}{}^{ab}  + \Lambda_{\mu}{}^{ab}\,,
\end{equation}
where $\omega_{\mu}{}^{ab}$ corresponds to the torsion-free Levi-Civita connection and can be written completely in terms of the tetrads, co-tetrads, and their derivatives, while $\Lambda_{\mu}{}^{ab}\,$ is an independent field called \textit{contorsion}. Using Eq.~\eqref{tetrad-postulate}, we can express the Ricci scalar through the spin connection, 
\begin{align}\label{Ricci}
	R(\Gamma) &= F_{\mu\nu}{}^{ab} e^\mu_a e^\nu_b\,,  \\
{\mathrm{ where}} \qquad \qquad F_{\mu\nu}{}^{ab} &= \partial_\mu A_\nu{}^{ab} - \partial_\nu A_\mu{}^{ab} + A_{\mu}{}^{a}{}_{c} A_{\nu}{}^{cb} -  A_{\nu}{}^{a}{}_{c} A_{\mu}{}^{cb}\,.
\end{align}
Then the action of gravity plus several fermion fields can be written as
\begin{align}
	S =& \int |e| d^4x \left(\frac{1}{2\kappa} F_{\mu\nu}{}^{ab}(A) e^\mu_a e^\nu_b +  \frac{1}{4} \sum_{i}\left(i\bar{\psi}_{i} \slashed{D}\psi_{i} + \text{h.c.}\right) - \frac{1}{2}\sum_{i}m_{i}\bar{\psi}_{i}\psi_{i} \right) \, \nonumber \\
    =& \int |e| d^4x \frac{1}{2\kappa} \left( F_{\mu\nu}{}^{ab}(\omega) + \hat{D}_{\mu}\Lambda_{\nu}{}^{ab} - \hat{D}_{\nu}\Lambda_{\mu}{}^{ab} + \Lambda_{\mu}{}^{a}{}_{c}\Lambda_{\nu}{}^{cb} - \Lambda_{\nu}{}^{a}{}_{c}\Lambda_{\nu}{}^{cb}\right)e^\mu_a e^\nu_b \nonumber \\
    & + \int |e| d^{4}x \sum_{i} \left(\frac{1}{4} \left( i\bar{\psi}_{i} \slashed{\partial}\psi_{i} + \text{h.c.}\right) + \frac{1}{8} e^{\mu c}\epsilon_{abcd}(\omega_{\mu}{}^{ab} + \Lambda_{\mu}{}^{ab})\bar{\psi}_{i}\gamma^{d}\gamma^{5}\psi_{i} - \frac{1}{2}\sum_{i}m_{i}\bar{\psi}_{i}\psi_{i} \right)\,, \label{action.1}
\end{align}
where $\kappa = 8\pi G$\,, and we have written $\hat{D}_\mu$ for the covariant derivative with respect to  $\omega_\mu.$ We will use the following notation: $\psi$ will denote a single field and $\Psi$ will denote a multiplet of fields; Greek indices $\alpha, \beta, \cdots$ of $\psi$ will denote fermions of definite flavor, Latin indices $i,j, \cdots$ will denote fermions of definite mass. From now on, we will refer to the $\psi_{i}$ as definite mass fields and the $\psi_{\alpha}$ will be called flavor fields. Since $\omega_\mu{}^{ab}$ and $\Lambda_\mu{}^{ab}$ are independent fields, it is in principle possible for their couplings to fermions to be diagonal in different bases. Of course, the coupling of $\omega_\mu{}^{ab}$ is diagonal in the definite mass basis. Latin indices $A, B, \cdots$ of $\psi$ will denote fermion fields such that the coupling with $\Lambda_{\mu}{}^{ab}$ is diagonal, and we will refer to $\psi_{A}$ as torsion fields. Of course, it is possible to go from one basis to another by a unitary transformation.  

The sum over $i,~\alpha$ and $A$ runs from $1$ to $(N_{a} + N)$, where $N_{a}$ is the number of active neutrinos and $N$ is the number of sterile neutrinos. In the SM, the $W, Z$ bosons couple to neutrinos by the terms
\begin{align}
    \La_{\text{CC}} =& -  \frac{g}{\sqrt{2}}\bar{\nu}_{f}\gamma^{\mu}\frac{1 - \gamma^{5}}{2}f W^{+}_{\mu}\,, \label{eq:la_cc} \\
    \La_{\text{NC}} =& - \frac{g}{\cos \theta_{W}}\bar{\nu}_{f}\gamma^{\mu}\frac{1 - \gamma^{5}}{2}\nu_{f} Z_{\mu}\,, \label{eq:la_nc}
\end{align}
where $g_{V} = T_{3}^{f} - 2 Q^{f} \sin^{2} \theta_{W}$ and $g_{A} = T_{3}^{f}$. Here $T_{3}^{f}$ is the third component of the weak isospin of the fermion and $Q^{f}$ is the electric charge of the same. In the two above Lagrangian $f = e, \mu, \tau$.
The equation of motion for $\Lambda_{\mu}{}^{ab}$ as obtained from Eq.~\eqref{action.1} is
\begin{align}
    \label{spin-connection}
	\Lambda_{\mu}{}^{ab} =& \frac{\kappa}{16}e^c_\mu\,\sum_{i}\bar{\psi}_{i}[\gamma_c,\sigma^{ab} ]_{+}\psi_{i} \equiv \frac{\kappa}{8}\epsilon^{abcd} e_{\mu c}\,\sum_{i}\bar{\psi}_{i} \gamma_d\gamma^5 \psi_{i} \,
.\end{align}
%
Let us discuss this contorsion field and its coupling to fermions in a little more detail. Since this is an algebraic equation of motion, we see that the contorsion field $\Lambda_{\mu}{}^{ab}$ does not propagate and is fully expressible in terms of the fermion fields. Furthermore, the spin connection acts as a gauge field of the local Lorentz symmetry and since left- and right-chiral fermions are in separate representations of the Lorentz group, they couple independently to $\Lambda_\mu{}^{ab}\,.$ Thus in general, the Lagrangian takes the form
%
%
\begin{align}
	\La_\psi &= \sum\limits_{i}	\left(\frac{i}{4}\bar{\psi}_{i}\gamma^\mu\partial_\mu\psi_{i} - 
	\frac{i}{4}\partial_\mu\bar{\psi}_{i}\gamma^\mu\psi_{i} + \frac{1}{8} e^{\mu c}\epsilon_{abcd}\omega_{\mu}{}^{ab}\bar{\psi}_{i}\gamma^{d}\gamma^{5}\psi_{i}\, - \frac{1}{2}m_{i}\bar\psi_{i}\psi_{i}\, \right) \notag \\ 
	&\qquad \qquad 
	+ \frac{1}{8} \Lambda_{\mu}{}^{ab} e^{\mu c}e_{abcd}
	\sum_{A} \left(-\lambda_A^{L}\bar{\psi}_{A\Le} \gamma^{d}\gamma^{5}\psi_{A\Le} + \lambda_A^{R}\bar{\psi}_{A\Ri} \gamma^{d}\gamma^{5}\psi_{A\Ri}\right)
	\,,\label{L_psi_all}
\end{align}
where the sum runs over all distinct fermion fields. The torsionless part of the covariant derivative is universal, so the corresponding term can be rotated back to the mass basis. However, due to the presence of the non-universal coupling, the same cannot be done for the $\Lambda_{\mu}{}^{ab}$ part of the connection. The generic solution for $\Lambda$ is therefore
\begin{equation}\label{chiral.torsion}
	\Lambda_{\mu}{}^{ab} = \frac{\kappa}{8}\epsilon^{abcd}e_{\mu c} \sum\limits_A \left(\lambda_{A}^{L}\bar{\psi}_{A\Le}\gamma_d \psi_{A\Le} + \lambda_{A}^R\bar{\psi}_{A\Ri}\gamma_d \psi_{A\Ri}\right)\,.
\end{equation}
Since the corresponding connection components $e^a_\nu e^b_\lambda\Lambda_{\mu}{}^{ab}$ are totally antisymmetric, the geodesic equation is unaffected, and all particles fall at the same rate. This interaction term is formally independent of the background metric, but of course it affects the metric through its contribution to the energy-momentum tensor. However, the resulting curvature will in general be small enough that the background can be taken to be approximately flat for the purpose of quantum field theory calculations. We emphasize that this term is not a ``curvature effect," nor some kind of ``modified gravity," but describes how ordinary fermions behave due to the fact that spacetime is fundamentally non-flat in nature. 

Since the contorsion field has an algebraic equation of motion and is thus non-propagating, we can substitute its solution back into the action and get
\begin{align}
    \La_\psi &= \sum\limits_{i}	\left(\frac{i}{4}\bar{\psi}_{i}\gamma^\mu\partial_\mu\psi_{i} - 
	\frac{i}{4}\partial_\mu\bar{\psi}_{i}\gamma^\mu\psi_{i} + \frac{1}{8} \omega_{\mu}{}^{ab} e_{abcd} e^{\mu c}  \,
	\bar{\psi}_{i}\gamma^{d}\gamma^{5} \psi_{i}\, - \frac{1}{2}m_{i}\bar\psi_{i}\psi_{i}\, \right) \notag \\ 
	&\qquad \qquad
	- \frac{3\kappa}{32} \left[\sum_{A}\left( \lambda_A^{L}\bar{\psi}_{A\Le} \gamma^{d}\psi_{A\Le} + \lambda_A^{R}\bar{\psi}_{A\Ri} \gamma^{d}  \psi_{A\Ri} \right) \right]^{2}\, \label{eq:effective_L}
.\end{align}

The interaction term is written in the basis (or fields) in which the spin torsion interaction is diagonal. We will avoid flavor oscillation of charged leptons, so  for the charged leptons the fields can be taken in the mass basis. 
We will also suppress the torsional interaction of the hadrons in this paper except with neutrinos.
The contorsion field has now disappeared from the Lagrangian, leaving behind a four-fermion interaction with unknown coupling constants.

This Lagrangian in Eq.~\eqref{eq:effective_L} describes fermions propagating in curved spacetime. However, for the purposes of neutrino physics, the Levi-Civita connection in the Lagrangian of Eq.~\eqref{eq:effective_L} is negligibly small and can be ignored. The quartic interaction obtained from integrating out the torsion is not necessarily very small -- may even be almost as large as weak interactions -- and therefore should not be neglected. The contribution of this interaction will be proportional to fermion densities and will be present for neutrinos going through the Earth. Then we can perform the QFT calculations with this new quartic interaction in the familiar QFT formalism defined on a flat manifold.


Let us consider the Lagrangian for the dynamics of neutrinos --- both active and sterile --- interacting with leptons and hadrons in the background, whose dynamics we will ignore for this paper. The torsional four-fermion interaction is given by 
\begin{align}
    \La_{\text{int}} = - \sum_{b} \left( \lambda_b^{L}\bar{\psi}_{b\Le} \gamma^{d}\psi_{b\Le} + \lambda_b^{R}\bar{\psi}_{b\Ri} \gamma^{d}  \psi_{b\Ri} \right) \sum_{A} \left( \lambda_A^{L}\bar{\psi}_{A\Le} \gamma_{d}\psi_{A\Le} + \lambda_A^{R}\bar{\psi}_{A\Ri} \gamma_{d}  \psi_{A\Ri} \right)\,.
\end{align}
In the above expression the subscript $b$ denotes fermions which are part of the background. Furthermore, we have absorbed a factor of $\sqrt{\tfrac{3\kappa}{16}}$ in all $\lambda$s. In the rest of the article we will be talking about neutrinos in the terrestrial environment, so neutrino density in the background is negligible. For neutrinos propagating through the 
Earth, the index $b$ runs over $u$ quarks, $d$ quarks, and electrons. For charged leptons and quarks the torsion basis (fields) will be taken as the same as the mass basis, as we mentioned earlier.

The presence of $\sqrt{\kappa}$ may lead one to think that the coupling constants $\lambda$ are of the order of $M_P^{-1}\,.$ This is a red herring. The contorsion field is independent of the metric (or the tetrad) and the coupling constants of these two types of fields are unrelated. Furthermore,  while $\kappa$ arises in gravitation because any quantum theory of gravity must reduce to General Relativity in the $\hbar \to 0$ limit, the coupling of torsion to fermion spin is a purely quantum effect. There is no classical limit of the torsional interaction between fermions and thus no natural scale. When we are doing flat space perturbation theory for this interaction, we need to ensure that the expectation values of the new interaction Hamiltonian is smaller than the typical eigenvalues of the free Hamiltonian. Looking at neutral current interactions, in particular neutrino oscillations in matter, we see that the coupling constants can be near the same size as the Fermi constant, or perhaps an order of magnitude smaller, for perturbation theory to work well for the energies and matter densities considered in usual experiments~\cite{Barick:2023yrs, Ghose:2025tgc, Barick:2025iwn}.


Further simplifications of the interaction Lagrangian can be made for dilute backgrounds. In that case, the background current can be replaced by its average value, which can be calculated in a covariant manner~\cite{Ghose:2023ttq, Pal:1989xs}, 
\begin{align}
    \La_{\text{int}} =& -  \left(\frac{1}{2}\sum_{b} n_{b}(\lambda_{b}^{L} + \lambda_{b}^{R})\right)
    \sum_{A} \left( \lambda_A^{L}\bar{\psi}_{A\Le} \gamma_{0}\psi_{A\Le} + \lambda_A^{R}\bar{\psi}_{A\Ri} \gamma_{0}  \psi_{A\Ri} \right) = -  \tilde{n} \sum_{A} \left( \lambda_A^{L}\bar{\psi}_{A\Le} \gamma_{0}\psi_{A\Le} + \lambda_A^{R}\bar{\psi}_{A\Ri} \gamma_{0}  \psi_{A\Ri} \right)
    \,. \label{eq:spin-torsion_effective_L}
\end{align}
{where $\tilde{n} = \frac{1}{2}\sum_{b} n_{b}(\lambda_{b}^{L} + \lambda_{b}^{R})$ is the $\lambda$-weighted density of the background fermions. }

Fermion dynamics in the presence of chiral torsion was discussed previously in~\cite{Barick:2023yrs, Barick:2025iwn}. However, there is a fundamental difference between the dynamics considered before and the dynamics considered now. In the previous works we started by assuming $\psi_{A} = \psi_{i}$. In this paper, we do not assume that the spin-torsion basis and the definite mass basis are the same, but just as the weak basis and the mass basis are related through a mixing, the torsion basis and the mass basis are related through a mixing as well.


\section{Hamiltonian for neutrino flavor oscillations}
\label{sec:qft}
%
In this section we derive the Hamiltonian for single neutrino states of definite flavor and definite helicity~\cite{Mohapatra:2004zh, Smirnov:2025wax, Akhmedov:2025enm}. In the Sec.~\ref{sec:seesaw} we have seen that the Type I SeeSaw mechanism generates Majorana type neutrinos. 
The free part of the Lagrangian is
%
\begin{align}
    \La_{\text{free}} = \sum_{i = 1}^{N_{a} + N}  \left(\frac{i}{4}(\bar{\nu}_{i}\gamma^{\mu}\partial_{\mu}\nu_{i} - \partial_{\mu}\bar{\nu}_{i}
    \gamma^{\mu}\nu_{i}) - \frac{1}{2}m_{i}\bar{\nu}_{i}\nu_{i}\right)
.\end{align}
Here $\nu_{i} = \nu_{i\Le} + \nu^{c}_{i\Le}$ is a full Majorana field and is given by the sum of the left-chiral Majorana fields and its charge conjugates. The definite mass fields can be expanded in terms of spinors of definite helicity $h$ as
%
\begin{align}
    \nu_{i} &= \int \frac{d^{3}p}{(2\pi)^{3/2}\sqrt{2E_{p}}} \sum_{h = \pm 1} (a^{(h)}_{i}(p)u^{(h)}_{i}(p)e^{-ip \cdot x} + a^{(h)^{\dagger}}_{i}(p)v^{(h)}_{i}(p)e^{i p \cdot x})\,, \nonumber \\
    \bar{\nu}_{i} &= \int \frac{d^{3}p}{(2\pi)^{3/2}\sqrt{2E_{p}}} \sum_{h = \pm 1} (a^{(h)^{\dagger}}_{i}(p)\bar{u}^{(h)}_{i}(p)e^{ip \cdot x} + a^{(h)}_{i}(p)\bar{v}^{(h)}_{i}(p)e^{-i p \cdot x})\,.
\end{align}
%
%
{The free Hamiltonian elements are}
\begin{align}
    (H_{0})^{hh^{\prime}}_{ij} = E_{i}\delta_{ij}\delta_{h h^{\prime}}\, \label{eq:vacuum_hamiltonian}
.\end{align}
Propagation of neutrinos through matter gives rise to effective Hamiltonian densities due to electroweak charged-current and neutral-current interactions~\cite{Wolfenstein:1977ue, Mikheyev:1985zog, Giunti:2007ry, Mohapatra:1998rq} 
\begin{align}
    \Ha^{\text{SM,CC}}_{\text{eff}} = \half{V_{CC}}\bar{\nu}_{e}\gamma^{0}(1-\gamma^{5})\nu_{e}\,,     \qquad     \Ha^{\text{SM,NC}}_{\text{eff}} = \half {V_{NC}}\sum_{\alpha = e, \mu, \tau}\bar{\nu}_{\alpha}\gamma^{0}(1-\gamma^{5})\nu_{\alpha}\,,
\end{align}
where $V_{CC} = \sqrt{2}G_{F}n_{e}$ and $V_{NC} = -\sqrt{2}G_{F}\frac{n_{n}}{2}$ as usual. The elements of the effective Hamiltonian due to the SM interactions in the ultrarelativistic limit can be calculated following~\cite{Giunti:2007ry}. They are given by%
\begin{align}
   (H^{\text{SM}}_{\text{eff}})^{h h^{\prime}}_{\alpha \alpha} =& (V_{M})_{\alpha \alpha}\delta_{h h^{\prime}}\begin{cases}1 \quad &\text{for} \quad h = -1\\ \frac{m^{2}_{\alpha}}{2E} \quad &\text{for} \quad h = +1\end{cases} \label{eq:sm_hamiltonian}
,\end{align}
where $(V_{M})_{\alpha \alpha} = (V_{CC}\delta_{\alpha e} + V_{NC})$\,, and $m_{\alpha}$ is the mass of the neutrino of flavor $\alpha$ if mixing is ignored. Although unphysical, the mass $m_{\alpha}$ shows that the contribution of $h = +1$ is heavily suppressed in the ultrarelativistic regime  \cite{Giunti:2003dg}.
The effective torsional four-fermion interaction Hamiltonian density for neutrinos due to the Lagrangian density in Eq.~\eqref{eq:spin-torsion_effective_L} is given by
\begin{align}
    \Ha^{\text{ST}}_{\text{eff}} = & 
    \frac{1}{2}{\tilde{n}} \sum_{A} \bar{\nu}_{A}(\lambda^{L}_{A}\gamma^{0}(1-\gamma^{5})+\lambda^{R}_{A}\gamma^{0}(1+\gamma^{5}))\nu_{A}\,\label{torsion-H}
.\end{align}
{As discussed before, the torsion fields $\nu_{A}$ are mixtures of the definite mass fields $\nu_{i}$, related by a {torsion-mass mixing matrix} $G$ such that }
\begin{equation}
    \Psi_{A} = G^{\dagger}\Psi_{m} = G^{\dagger}\tilde{U}^{\dagger}\Psi\,.
\end{equation}
We will denote the multiplet of definite mass fields by the subscript $m$\,, the multiplet of torsion fields by the subscript $A$\,, and the multiplet of flavor fields with no subscript. 
We know that when two multiplets formed from the same set of fields are related by a mixing matrix, the corresponding states are related by~\cite{Bilenky:1987ty}
\begin{equation}
    \ket{\nu^{(h)}_{A}} = \sum_{i}(G^{T})_{Ai}\ket{\nu^{(h)}_{i}} = \sum_{i} G_{iA} \ket{\nu^{(h)}_{i}} = \sum_{\alpha} (\tilde{U}G)_{\alpha A} \ket{\nu^{(h)}_{\alpha}}\,. \label{eq:G_matrix}
\end{equation}
{The exact nature of the torsion-mass mixing matrix $G$ is dependent on the nature of the torsion states. If the torsion states are Dirac then the $G$ is complex and if the torsion states are Majorana in nature then the matrix $G$ is real.} Following arguments similar to the case for electroweak interactions~\cite{Giunti:2007ry} we can calculate the elements of the effective Hamiltonian in the ultrarelativistc limit to be
\begin{align}
          &(H_{\text{eff}}^{\text{ST}})^{hh^{\prime}}_{AA} = \tilde{n}\delta_{h h^{\prime}}\begin{cases}
         \left(\lambda_{A}^{L} + \frac{m^{2}_{A}}{2 E}\lambda_{A}^{R}\right) \quad &\text{for} \quad h = -1 \\
         \left(\lambda_{A}^{L}\frac{m^{2}_{A}}{2 E} + \lambda_{A}^{R}\right) \quad &\text{for} \quad h =+1
     \end{cases}\, \label{eq:spin-torsion_hamiltonian_dirac}, \\
     &\text{when the torsion fields are Dirac, and} \nonumber \\  
          &(H_{\text{eff}}^{\text{ST}})^{hh^{\prime}}_{AA} = \tilde{n}\delta_{h h^{\prime}}\begin{cases}
         (\lambda_{A}^{L} - \lambda_{A}^{R}) \quad &\text{for} \quad h = -1 \\
         (-\lambda_{A}^{L} + \lambda_{A}^{R}) \quad &\text{for} \quad h =+1
     \end{cases}\,,\label{eq:spin-torsion_hamiltonian_majorana} \\
     &\text{when the torsion fields are Majorana,} \notag
     \end{align}
where $m_{A}$ is the mass of neutrino $\nu_{A}$ if torsion-mass mixing is ignored. {For more details of how this Hamiltonian is calculated, see App.~\ref{app:matter_pot}.}

The Hamiltonian elements in Eq.~\eqref{eq:spin-torsion_hamiltonian_dirac} were calculated for neutrino states. The Hamiltonian elements calculated for antineutrino states will gain a relative $(-1)$. However, for Majorana-like torsion states, no such change will be there --- neutrino states and antineutrino states are identical. This is a reflection of the fact that Majorana fermions have half the degrees of freedom than that of Dirac fermions.

Neutrino flavor states evolve according to the Schr\"{o}dinger equation
\begin{align}
    i\frac{d}{dt}\ket{\Psi (t)} = H\ket{\Psi(t)} = (H_{0} + H^{\text{SM}}_{\text{eff}} + H^{\text{ST}}_{\text{eff}})\ket{\Psi (t)}\,. 
\end{align}
The general flavor state $\ket{\Psi(t)}$ is a vector in the Hilbert space labeled by helicity $\otimes$ flavor. That means we can use the completeness relation $\sum_{\alpha,h} \ket{\nu^{(h)}_{\alpha}}\bra{\nu^{(h)}_{\alpha}} = \mathbb{I}$. Expanding a neutrino state as $\ket{\Psi(t)} = \sum_{\alpha, h} c^{(h)}_{\alpha} \ket{{\nu^{(h)}_{\alpha}}}$\,, we can write
\begin{align}
    i\frac{d}{d t} c^{h}_{\alpha} = \sum_{\beta, h^{\prime}}H^{hh^{\prime}}_{\alpha \beta} c^{h^{\prime}}_{\beta}\,. \label{eq:schro_full}
\end{align}
The vacuum Hamiltonian in Eq.~\eqref{eq:vacuum_hamiltonian} contributes equally to both helicities. Hence we see that there is no chirality flip in vacuum, which was also noted in~\cite{Smirnov:2025wax, Akhmedov:2025enm}. In presence of matter the two helicities receive unequal contributions 
and that results in a non-zero probability of chirality flip. However, this probability of chirality change is heavily suppressed in the ultrarelativistic limit. As the neutrinos detected or produced in all known experiments are ultrarelativistic, the chirality flip can still be ignored. This argument is elaborated in Appendix~\ref{sec:helicity}. We note that there is an independent line of investigation which concludes that there must be vacuum chiral oscillations \cite{Blasone:2025hjw, Bittencourt:2024yxi, Kimura:2021qlh, Ge:2020aen}.

Let us consider Eq.~\eqref{eq:schro_full} for Dirac-type torsion states. Suppose we are working with 2 active neutrinos and 1 sterile neutrino. Following the arguments in Appendix~\ref{sec:helicity}, the two helicity sectors can be decoupled, and we can write the flavor evolution in the negative helicity sector as
\begin{align}
    i\frac{d}{d t} \begin{pmatrix}c^{(-)}_{e}\\ c^{(-)}_{\mu}\\ c^{(-)}_{s}\end{pmatrix} = \left[\tilde{U} \begin{pmatrix}E_{0} & 0 & 0\\ 0 & E_{1} & 0\\ 0 & 0 & E_{2}\end{pmatrix} \tilde{U}^{\dagger} + \begin{pmatrix} V_{CC}+V_{NC}& 0 & 0\\ 0 & V_{NC} & 0\\ 0 & 0 & 0\end{pmatrix} + \tilde{n}(\tilde{U}G)\begin{pmatrix}\lambda^{L}_{1} & 0 & 0\\ 0 & \lambda_{2}^{L} & 0\\ 0 & 0 & \lambda_{3}^{L}\end{pmatrix}(\tilde{U}G)^{\dagger}\right] \begin{pmatrix}c^{(-)}_{e}\\ c^{(-)}_{\mu}\\ c^{(-)}_{s}\end{pmatrix}.
    \label{eq:flavor_eqn_neg_h_dirac}
\end{align}
We can see from Eq.~\eqref{eq:spin-torsion_hamiltonian_dirac} that the matter potential also has a contribution containing $\lambda^{R}$, however it is proportional to a term $\dfrac{m^{2}_{A}}{2 E}$ and hence is suppressed. If the torsion states are Majorana in nature, then the matter potential due to the Spin-Torsion interaction is given by Eq. \eqref{eq:spin-torsion_hamiltonian_majorana}. The corresponding Schr\"{o}dinger equation is
\begin{align}
    i\frac{d}{d t} \begin{pmatrix}c^{(-)}_{e}\\ c^{(-)}_{\mu}\\ c^{(-)}_{s}\end{pmatrix} =& \left[\tilde{U} \begin{pmatrix}E_{0} & 0 & 0\\ 0 & E_{1} & 0\\ 0 & 0 & E_{2}\end{pmatrix} \tilde{U}^{\dagger} + \begin{pmatrix} V_{CC}+V_{NC}& 0 & 0\\ 0 & V_{NC} & 0\\ 0 & 0 & 0\end{pmatrix}\right. \notag \\
    & \qquad \left.+ \tilde{n}(\tilde{U}G)\begin{pmatrix}\lambda^{L}_{1} - \lambda^{R}_{1} & 0 & 0\\ 0 & \lambda_{2}^{L} - \lambda_{2}^{R} & 0\\ 0 & 0 & \lambda_{3}^{L} - \lambda_{3}^{R}\end{pmatrix}(\tilde{U}G)^{\dagger}\right] \begin{pmatrix}c^{(-)}_{e}\\ c^{(-)}_{\mu}\\ c^{(-)}_{s}\end{pmatrix}
    \label{eq:flavor_eqn_neg_h}
.\end{align}
In the rest of the article, we will focus on the case of Majorana-type torsion states as seen in Eq. \eqref{eq:flavor_eqn_neg_h}. We will also use $\lambda_{A}$ to denote $\lambda_{A}^{L} - \lambda_{A}^{R}$. Looking at Eqs.~\eqref{eq:flavor_eqn_neg_h_dirac} and \eqref{eq:flavor_eqn_neg_h}, we see that putting $\lambda^{R} = 0$ in the flavor evolution equation for the Majorana-like torsion states in Eq.~\eqref{eq:flavor_eqn_neg_h} results in the flavor evolution equation of the Dirac-like torsion states in Eq.~\eqref{eq:flavor_eqn_neg_h_dirac}. Even though we consider only the case of Majorana-type torsion states, we still capture all the necessary physics. 

\section{1 active and 1 sterile neutrino}
\label{sec:1p1}

To understand the effects of the Spin-Torsion interaction on Majorana neutrinos, we start with a simple $1+1$ scenario. In this scenario, we start with one active neutrino which we will call $\nu_e$ and one sterile neutrino which we will call $N_s$\,. The mass matrix in this case is a 2x2 complex symmetric matrix. The mass Lagrangian is then given by
\begin{align}
    \La_{\text{mass}} = - \frac{1}{2}\begin{pmatrix}\bar{\nu}_{e} & \bar{N}^{c}_{s}\end{pmatrix}\begin{pmatrix}0 & m_{e s}e^{i \phi}\\ m_{e s}e^{i \phi} & M\end{pmatrix}\begin{pmatrix}\nu^{c}_{e}\\ N_{s}\end{pmatrix} + \hc\,,
\end{align}
with real $(m_{e s}, \phi, M)$ . The mass eigenvalues are $(\mu, M_{1}) = (m^{2}_{e s}/M, M + m^{2}_{e s}/M)$\,, so that the mass-squared difference 
is~\cite{Giunti:2007ry}
\begin{align}
    \Delta m^{2} = M^{2}_{1} - \mu^{2} = \left(M + \frac{m^{2}_{e s}}{M}\right)^{2} - \frac{m^{4}_{e s }}{M^{2}} = M^{2} + 2 m^{2}_{e s}\,.
\end{align}
Following our discussion in Sec.~\ref{sec:seesaw}, we can write the {flavor-mass mixing matrix} to be
\begin{align}
    \tilde{U} = \begin{pmatrix}\left(1 - \frac{1}{2}\frac{m^{2}_{e s}}{M^{2}}\right)ie^{i (\beta + \phi)} & \frac{m_{e s}}{M}e^{i \phi}\\ -i\frac{m_{e s}}{M}e^{i \beta} & 1 - \frac{1}{2}\frac{m^{2}_{e s}}{M^{2}}\,\end{pmatrix} = \begin{pmatrix}e^{i \phi} & 0\\ 0 & 1\end{pmatrix}\begin{pmatrix}\left(1 - \frac{1}{2}\frac{m^{2}_{e s}}{M^{2}}\right) & \frac{m_{e s}}{M}\\ - \frac{m_{e s}}{M} & \left(1 - \frac{1}{2}\frac{m^{2}_{e s}}{M^{2}}\right)\end{pmatrix}\begin{pmatrix}i e^{i \beta} & 0\\ 0 & 1\end{pmatrix}\,.
\end{align}
The Majorana phase $\beta$ cannot be removed by redefining the fields. On the other hand, the phase $\phi$ can be absorbed into the definition of $\nu_{e}$. Hence, the angle $\phi$ is not a physical angle and cannot appear in the effective mass squared difference in matter in the presence of both SM and Spin-Torsion interactions. This redefinition of Dirac neutrino fields is equivalent to starting with real $M_{D}$ and $M_{M}$ in the mass matrix with complex Majorana mass term for active neutrinos~\cite{Giunti:2007ry}. However, we are explicitly starting with a framework where the active neutrinos do not have Majorana masses to begin with. Hence, we will proceed with complex $M_{D}$ and the entries of the Hamiltonian will contain $\phi$. The {flavor-mass mixing matrix} diagonalizes the mass Lagrangian as
\begin{align}
    \La_{\text{mass}} =& -\frac{1}{2}\begin{pmatrix}\bar{\nu}_{e} & \bar{N}^{c}_{s}\end{pmatrix}\tilde{U}\begin{pmatrix}\mu & 0\\ 0 & M_{1}\end{pmatrix}\tilde{U}^{T}\begin{pmatrix}\nu^{c}_{e}\\ N_{s}\end{pmatrix} + \hc = -\frac{1}{2}\begin{pmatrix}\bar{\nu}_{1\Le} & \bar{\nu}_{2\Le}\end{pmatrix}\begin{pmatrix}\mu & 0\\ 0 & M_{1}\end{pmatrix}\begin{pmatrix}\nu^{c}_{1\Le}\\ \nu^{c}_{2\Le}\end{pmatrix} + \hc\,.
\end{align}
The two neutrinos resulting from the diagonalisation are $\nu_{1}$ with mass $\mu$ and $\nu_{2}$ with mass $M_{1}$. Of these, $\nu_{1\Le}$ mostly contains the active field $\nu_{e}$\,, while $\nu_{2\Le}$ is mostly the sterile neutrino $N^c_{s}$. The flavor evolution equation is
\begin{align}
    i \frac{d}{dt}\begin{pmatrix}c^{(-)}_{e}\\ c^{(-)}_{s}\end{pmatrix} = E_{0} + \biggl[ \tilde{U}\begin{pmatrix}-\frac{M^{2}_{1}}{4 p} + \frac{\mu^{2}}{4 p} & 0\\ 0 & \frac{M^{2}_{1}}{4 p} - \frac{\mu^{2}}{4 p}\end{pmatrix} \tilde{U}^{\dagger} + \begin{pmatrix}\frac{V_{SM}}{2} & 0\\ 0 & - \frac{V_{SM}}{2}\end{pmatrix} + \frac{(\tilde{n}\lambda_{2}-\tilde{n}\lambda_{1})}{2}(\tilde{U} G) \begin{pmatrix} -1 & 0\\ 0 & 1 \end{pmatrix}(\tilde{U}G)^{\dagger}\biggr]\begin{pmatrix}c^{(-)}_{e}\\ c^{(-)}_{s}\end{pmatrix}\,. \label{eq:schro_1p1}
\end{align}
Here we have written $E_{0} = p + \dfrac{M^{2}_{1} + \mu^{2}}{4 p} + \dfrac{V_{SM}}{2} + \dfrac{\tilde{n}\lambda_{1} + \tilde{n}\lambda_{2}}{2}$, and also written the matter potential due to electroweak interactions as $V_{SM} = V_{CC} + V_{NC}$. As discussed earlier in Eq.~\eqref{eq:G_matrix}, the torsion fields are related to the definite mass fields through a torsion-mass mixing matrix $G$. The exact form of $G$ is dependent on the nature of the fields $\nu_{A}$ --- when the $\nu_{A}$s are Majorana fields, $G$ is real and has only one free parameter in the $1+1$ scenario. Then the {torsion-mass mixing matrix is given by} 
\begin{align}
    G = \begin{pmatrix}\cos \theta & \sin \theta\\ - \sin \theta & \cos \theta\end{pmatrix}\,.
\end{align} 
Then the Hamiltonian in Eq.~\eqref{eq:schro_1p1} can be calculated to be
\begin{align}
    i\frac{d}{dt}\begin{pmatrix}c^{(-)}_{e}\\ c^{(-)}_{s}\end{pmatrix} = H\begin{pmatrix}c^{(-)}_{e}\\ c^{(-)}_{s}\end{pmatrix} = \biggl[E_{0} + \begin{pmatrix}H_{11} & H_{12}\\ H^{*}_{12} & - H_{11}\end{pmatrix}\biggr] \begin{pmatrix}c^{(-)}_{e}\\ c^{(-)}_{s}\end{pmatrix}\,,
\end{align}
where we have defined
\begin{align}
    H_{11} =& - \frac{M^{2}}{4 p} + \frac{V_{SM}}{2} - \left(\frac{1}{2}\cos2\theta  + \frac{\sin\beta 
    \sin2\theta m_{e s}}{M} - \frac{\cos 2 \theta m^{2}_{e s}}{M^{2}}\right)(\tilde{n}\lambda_{2}-\tilde{n}\lambda_{1}) \nonumber \\
    H_{12} =& e^{i \phi}\left(\frac{M m_{e s}}{2 p} + \frac{\sin 2\theta (2 \sin \beta m^{2}_{e s} + i M^{2} e^{i \beta}) + 2 \cos 2\theta M m_{e s}}{2 M^{2}}(\tilde{n}\lambda_{2} - \tilde{n}\lambda_{1})\right)\,.
\end{align}
In the presence of matter, the energy eigenvalues are
\begin{align}
    E^{\prime}_{1,2} = E_{0} \pm \sqrt{H^{2}_{11} + |H_{12}|^{2}}\,.
\end{align}
The effective mass squared difference in matter is given by 
\begin{align}
    \Delta \tilde{m}^{2} =& 2 p (E^{\prime}_{2} - E^{\prime}_{1})  \nonumber \\
    =& 4 p \left[\left(- \frac{M^{2}}{4 p} + \frac{V_{SM}}{2} - \left(\frac{1}{2}\cos 2 \theta  + \frac{\sin \beta \sin 2 \theta m_{e s}}{M} - \frac{\cos 2 \theta m^{2}_{e s}}{M^{2}}\right)(\tilde{n}\lambda_{2}-\tilde{n}\lambda_{1})\right)^{2}\right. \nonumber \\
    & \left.+ \left| \frac{M m_{e s}}{2 p} + \frac{\sin 2\theta (2 \sin \beta m^{2}_{e s} + i M^{2} e^{i\beta}) + 2 \cos 2\theta M m_{e s}}{2 M^{2}} (\tilde{n}\lambda_{2}-\tilde{n}\lambda_{1})\right|^{2}\right]^{1/2}. \label{eq:msd_1p1}
\end{align}
From the above expression we see that the effective mass squared difference is dependent on the {torsion-mass mixing angle} $\theta$ as well as on the Majorana phase $\beta$. However, the dependence on the Majorana phase is only present if the torsion fields are not aligned with the definite mass fields. If the torsion fields are aligned, then $\theta = 0$ and the effective mass squared difference does not depend on the Majorana phase. Also, the angle $\phi$ is absent from the effective mass squared difference even in the presence of torsion. This is expected as we have already seen that $\phi$ is not a physical phase. For antineutrinos, $\lambda_{A}, V_{SM}$, and the Majorana phase $\beta$ will all change sign. {This shows that the effective mass squared difference in matter for antineutrinos will be different from that of neutrinos in general. This is a consequence of the matter effect.}
%

In the high momentum limit, the effective mass squared difference is 
%
%
\begin{align}
    \Delta \tilde{m}^{2} = 4 p &\left[\left(+ \frac{V_{SM}}{2} - \left(\frac{1}{2}\cos 2 \theta + \frac{\sin \beta \sin 2 \theta m_{e s}}{M} + \frac{\cos 2 \theta m^{2}_{e s}}{M^{2}} \right) (\tilde{n}\lambda_{2}-\tilde{n}\lambda_{1})\right)^{2}\right. \nonumber \\
    & \left.+ \biggl\lvert \left(\frac{i \sin 2 \theta e^{i \beta}}{2} + \sin 2 \theta \sin \beta \frac{m^{2}_{e s}}{M^{2}} + \frac{m_{e s}}{M}\cos 2 \theta\right)(\tilde{n}\lambda_{2} - \tilde{n}\lambda_{1}) \biggr\rvert^{2}\right]^{1/2}.
\end{align}
Thus in the high momentum limit the effective mass squared difference is dominated by the matter effects through both the SM and Spin-Torsion interactions. However, if we take torsion fields to be aligned with the definite mass fields, i.e., if $\theta = 0$,  $\Delta \tilde{m}^{2}$ is the same in both the neutrino and the antineutrino sector. A non-zero $\theta$ breaks that symmetry in high momentum limit.

\section{The 2 active and 1 sterile neutrino scenario}
\label{sec:2p1}
In the previous section, we considered the simplified 1 active + 1 sterile neutrino paradigm. In this section, we will concentrate on the scenario of 2 active neutrinos $\nu_{e}, \nu_{\mu}$ and 1 sterile neutrino $N_{s}$. The active neutrinos are left-chiral and the sterile neutrinos are right-chiral. The corresponding mass Lagrangian in this case is
\begin{align}
    \La_{\text{mass}} = -\frac{1}{2}\begin{pmatrix}\bar{\nu}_{e} & \bar{\nu}_{\mu} & \bar{N}^{c}_{s} \end{pmatrix}\begin{pmatrix}0 & 0 & m_{e s}e^{i \phi_{1}}\\ 0 & 0 & m_{\mu s}e^{i \phi_{2}}\\ m_{e s}e^{i \phi_{1}} & m_{\mu s}e^{i \phi_{2}} & M\end{pmatrix}\begin{pmatrix}\nu^{c}_{e}\\ \nu^{c}_{\mu}\\ N_{s}\end{pmatrix} + \hc
\end{align}
Here $m_{e s}, m_{\mu s}, M, \phi_{1}, \phi_{2}$ are all real parameters, with $M\gg m_{es}, m_{\mu s}$. The mass eigenvalues are $(0, \mu, M_{1}) = (0, m^{2}/M, M + m^{2}/M)$, where  $m = \sqrt{m_{e s}^{2} + m_{\mu s}^{2}}$. Thus the mass spectrum has 1 massless neutrino, 1 light neutrino and 1 heavy neutrino. The flavor-mass mixing matrix is then
\begin{align}
    \tilde{U} =& \begin{pmatrix}e^{-i \phi_{2}}\frac{m_{\mu s}}{m} \quad & e^{i \phi_{1}}\frac{m_{e s}}{m}\left(1 - \frac{1}{2}\frac{m^{2}}{M^{2}}\right)\quad & \frac{m_{e s}}{M}e^{i \phi_{1}}\\ - e^{-i \phi_{1}}\frac{m_{e s}}{m}\quad & e^{i \phi_{2}}\frac{m_{\mu s}}{m}\left(1 - \frac{1}{2}\frac{m^{2}}{M^{2}}\right)\quad & \frac{m_{\mu s}}{M}e^{i \phi_{2}}\\ 0 & -\frac{m}{M} & 1\end{pmatrix}\begin{pmatrix}ie^{i\alpha} & 0 & 0\\ 0 & ie^{i\beta} & 0\\ 0 & 0 & 1\end{pmatrix} \nonumber \\
    =& \begin{pmatrix}e^{-i \phi_{2}} & 0 & 0\\ 0 & e^{-i\phi_{1}} & 0\\ 0 & 0 & 1\end{pmatrix}\begin{pmatrix}\frac{m_{\mu s}}{m} & e^{i \phi}\frac{m_{e s}}{m}\left(1 - \frac{1}{2}\frac{m^{2}}{M^{2}}\right) & e^{i \phi}\frac{m_{e s}}{M}\\ - \frac{m_{e s}}{m} & e^{i \phi}\frac{m_{\mu s}}{m}\left(1 - \frac{1}{2}\frac{m^{2}}{M^{2}}\right) & e^{i \phi}\frac{m_{\mu s}}{M}\\ 0 & -\frac{m}{M} & 1\end{pmatrix}\begin{pmatrix}ie^{i\alpha} & 0 & 0\\ 0 & ie^{i\beta} & 0\\ 0 & 0 & 1\end{pmatrix} \label{eq:mixing_matrix_flavor}
.\end{align}
In the second line, we have defined $\phi = \phi_{1} + \phi_{2}$. In this case, the flavor evolution equation turns out to be 
\begin{align}
    i\frac{d}{dt}\begin{pmatrix}c^{(-)}_{e}\\ c^{(-)}_{\mu}\\ c^{(-)}_{s}\end{pmatrix} = \left[p + \tilde{U}\begin{pmatrix}0 & 0 & 0\\ 0 & \frac{\mu^{2}}{2 p} & 0\\ 0 & 0 & \frac{M^{2}_{1}}{2 p}\end{pmatrix}\tilde{U}^{\dagger} + \begin{pmatrix}V_{CC} + V_{NC} & 0 & 0\\ 0 & V_{NC} & 0\\ 0 & 0 & 0\end{pmatrix} + (\tilde{U}G)\begin{pmatrix}\tilde{n}\lambda_{1} & 0 & 0\\ 0 & \tilde{n}\lambda_{2} & 0\\ 0 & 0 & \tilde{n}\lambda_{3}\end{pmatrix}(\tilde{U}G)^{\dagger}\right]\begin{pmatrix}c^{(-)}_{e}\\ c^{(-)}_{\mu}\\ c^{(-)}_{s}\end{pmatrix} \label{eq:schro_eqn_2p1}
.\end{align}

The {torsion-mass mixing matrix} $G$ is a real $3 \times 3$ mixing matrix and can be parametrized by three mixing angles in a manner analogous to the usual mass-flavor mixing matrix,
\begin{align}
    G = O_{23} O_{13} O_{12}\,.
\end{align}
Here $O_{ij}$ is the SO(3) matrix corresponding to rotation in the $(ij)$ plane, e.g., 
\begin{equation}
    O_{12} = \begin{pmatrix}\cos \phi_{12} & \sin \phi_{12} & 0\\ - \sin \phi_{12} & \cos \phi_{12}  & 0 \\ 0 & 0 & 1\end{pmatrix}\,,
\end{equation}
and similarly for the others. 
Finding the energy eigenvalues for a general $G$, even approximately, is quite difficult. So we will resort to some simplifications.

\subsection{Torsion-mass alignment}
\label{subsec:aligned_2p1}

{The simplest such scenario is when the torsion fields are completely aligned with the mass fields. Then $G = \mathbb{I}_{3}$.} Even when only SM interactions are present, the exact solution of 3 neutrino oscillation is already quite complicated~\cite{Kimura:2002hb, Barger:1980tf, Zaglauer:1988gz} and not always necessary. Since we are already working in the SeeSaw limit {$|M_{D}| \ll M_{M}$}, we can work out the approximate energy eigenvalues. This is done by a perturbative calculation, using the Hamiltonian elements shown in Appendix~\ref{app:2p1_aligned}.
Using Eq.~\eqref{eq:leading_order_Hprime_aligned_2p1}, we can calculate the energy eigenvalues correct up to $\mathcal{O}((m/M)^{0})$  to be
\begin{align}
    E^{\prime}_{1} =& p + V_{NC} + \frac{1}{2}V_{CC} + \frac{1}{2}(\tilde{n}\lambda_{2}+\tilde{n}\lambda_{1}) - \frac{1}{2}\mathcal{A}\,, \nonumber \\
    E^{\prime}_{2} =& p + V_{NC} + \frac{1}{2}V_{CC} + \frac{1}{2}(\tilde{n}\lambda_{2}+\tilde{n}\lambda_{1}) + \frac{1}{2}\mathcal{A}\,, \nonumber \\
    E^{\prime}_{3} =& p + \frac{M^{2}}{2 p} + \tilde{n}\lambda_{3}\,,
    \label{eq:leading_order_eigenval_align_2p1}
\end{align}
where we have defined 
\begin{align}
    \mathcal{A} = & \sqrt{\left(V_{CC} - \left(\frac{m^{2}_{\mu s}}{m^{2}} - \frac{m^{2}_{e s}}{m^{2}}\right)(\tilde{n}\lambda_{2}-\tilde{n}\lambda_{1})\right)^{2} + \left(\frac{2m_{e s}m_{\mu s}}{m^{2}}(\tilde{n}\lambda_{2}-\tilde{n}\lambda_{1})\right)^{2}}\,.
\end{align}
Eq.~\eqref{eq:leading_order_eigenval_align_2p1} gives us the modified dispersion relation in the presence of matter. The interaction with the background matter changes the effective mass of the neutrinos as it propagates through the matter. The effective mass squared differences in matter are 
\begin{align}
    \Delta \tilde{m}^{2}_{21} = 2p(E^{\prime}_{2} - E^{\prime}_{1}) =& 2p\mathcal{A}\,, \nonumber \\
    \Delta \tilde{m}^{2}_{31} = 2 p (E^{\prime}_{3} - E^{\prime}_{1}) =& M^{2} + 2 p \biggl[\tilde{n}\lambda_{3} - V_{NC} - \frac{1}{2}V_{CC} - \frac{1}{2}(\tilde{n}\lambda_{2}+\tilde{n}\lambda_{1}) + \frac{1}{2}\mathcal{A}\biggr]\,.
    \label{eq:msd_leading_order_align_2p1}
\end{align}
The tildes above the $m$ indicate that the effective mass squared differences are calculated in the presence of matter, including both electroweak interactions and torsional interactions. Further insights about the effective mass squared differences can be obtained by looking back at the {flavor-mass mixing matrix} $\tilde{U}$ of Eq. \eqref{eq:mixing_matrix_flavor}. We see that it can be expressed as
\begin{align}
    \tilde{U} = \begin{pmatrix}e^{-i \phi_{2}} & 0 & 0\\ 0 & e^{-i\phi_{1}} & 0\\ 0 & 0 & 1\end{pmatrix}\begin{pmatrix}1 - \frac{1}{2}\frac{m^{2}_{e s}}{M^{2}}e^{i\phi_{2}} & -\frac{m_{e s}m_{\mu s}}{2M^{2}}e^{i\phi_{1}} & \frac{m_{e s}}{M}e^{i \phi}\\ -\frac{m_{e s}m_{\mu s}}{2 M^{2}}e^{i\phi_{2}} & 1 - \frac{1}{2}\frac{m^{2}_{\mu s}}{M^{2}}e^{i\phi_{1}} & \frac{m_{\mu s}}{M}e^{i \phi}\\ -\frac{m_{e s}}{M}e^{-i \phi_{1}} & - \frac{m_{\mu s}}{M}e^{-i \phi_{2}} & 1 - \frac{1}{2}\frac{m^{2}}{M^{2}}\end{pmatrix}\begin{pmatrix}\frac{m_{\mu s}}{m} & \frac{m_{e s}}{m}e^{i\phi} & 0\\ -\frac{m_{e s}}{m} & \frac{m_{\mu s}}{m}e^{i\phi} & 0\\ 0 & 0 & 1\end{pmatrix}\begin{pmatrix}ie^{i\alpha} & 0 & 0\\ 0 & ie^{i\beta} & 0\\ 0 & 0 & 1\end{pmatrix}. \label{eq:flavor-mass-mixing-2p1}
\end{align}
%
This matrix contains two angles $\phi_{1, 2}$ which can be absorbed into the Dirac-type flavor fields. However, the combination $\phi_{1} + \phi_{2} = \phi$ of these two angles can not be redefined away.

The terms independent of $m/M$ in the flavor-mass mixing matrix $\tilde{U}$ in Eq.~\eqref{eq:flavor-mass-mixing-2p1} form a block in the active sector, which looks like a 2-neutrino mixing matrix
\begin{align}
    \begin{pmatrix}
        \cos \chi & \sin \chi \\
        -\sin \chi & \cos \chi
    \end{pmatrix}
,\end{align}
multiplied by a phase matrix, where $\cos \chi = \tfrac{m_{\mu s}}{m}$ and $\sin \chi = \tfrac{m_{e s}}{m}$. {Flavor mixing of two neutrinos in the presence of chiral torsion was studied in}~\cite{Ghose:2023ttq, Barick:2023yrs} assuming that the torsion fields are always parallel with the definite mass fields. The effective mass squared difference in the presence of matter in that case was found to be
\begin{align}
    \Delta \tilde{m}^{2} = 2 p \sqrt{\left(\frac{\Delta m^{2}}{2 p}\cos 2 \chi + (\tilde{n}\lambda_{2}-\tilde{n}\lambda_{1})\cos 2\chi - V_{CC}\right)^{2} + \left(\frac{\Delta m^{2}}{2 p} \sin 2\chi + (\tilde{n}\lambda_{2}-\tilde{n}\lambda_{1})\sin2\chi \right)^{2}}\,.
    \label{eq:effective_mass_sqaured_difference_ghose}
\end{align}
Using the values of $\cos \chi$ and $\sin \chi$ {mentioned above,} we can verify that the mass-squared difference $\Delta \tilde{m}^{2}_{21}$ is the same as the effective mass-squared difference in matter in Eq.~\eqref{eq:effective_mass_sqaured_difference_ghose}. {Hence, we see that the resultant effective mass squared differences in matter derived for neutrinos in Type I SeeSaw mechanism are consistent with the previous result in~\cite{Barick:2023yrs}}.

 Collecting all the terms of $\mathcal{O}(m/M)$ of the Hamiltonian in Eq.~\eqref{eq:first_order_Hprime_aligned_2p1}, we see that the eigenvalues do not get any correction in the first-order perturbation theory. The leading perturbative corrections come at the second order in $m/M$. Using the expressions in  Eqs.~\eqref{eq:first_order_Hprime_aligned_2p1} and \eqref{eq:second_order_Hprime_aligned_2p1}, we can calculate the energy eigenvalues correct up to $\mathcal{O}((m/M)^{2})$,
\begin{align}
    E^{\prime}_{1} =& p + V_{NC} + \frac{1}{2}V_{CC} + \frac{1}{2}(\tilde{n}\lambda_{2}+\tilde{n}\lambda_{1}) - \frac{1}{2}\mathcal{A} + \mathcal{A}_{1} + \frac{m^{2}_{e s}}{2 p} - \frac{m^{2}_{e s}}{M^{2}}(\tilde{n}\lambda_{2}-\tilde{n}\lambda_{3})\,, \nonumber \\
    E^{\prime}_{2} =& p + V_{NC} + \frac{1}{2}V_{CC} + \frac{1}{2}(\tilde{n}\lambda_{2}+\tilde{n}\lambda_{1}) + \frac{1}{2}\mathcal{A} + \mathcal{A}_{2} + \frac{m^{2}_{\mu s}}{2 p} - \frac{m^{2}_{\mu s}}{M^{2}}(\tilde{n}\lambda_{2} - \tilde{n}\lambda_{3})\,, \nonumber \\
    E^{\prime}_{3} =& p + \frac{M^{2}}{2 p} + \tilde{n}\lambda_{3} + \frac{m^{2}}{2 p} + \frac{m^{2}}{M^{2}}(\tilde{n}\lambda_{2} - \tilde{n}\lambda_{3}) - \mathcal{A}_{1} - \mathcal{A}_{2}\,. \label{eq:second_order_eigenval_align_2p1}
\end{align}
Here we have written
\begin{align}
    \mathcal{A}_{1} =& \frac{M^{2} m^{2}_{e s}}{4 p^{2}}\frac{\left(1 - \frac{2 p}{M^{2}}(\tilde{n}\lambda_{2}-\tilde{n}\lambda_{3})\right)^{2}}{V_{NC} + \frac{1}{2}V_{CC} + \frac{1}{2}(\tilde{n}\lambda_{1}+\tilde{n}\lambda_{2}) - \tilde{n}\lambda_{3} - \frac{1}{2}\mathcal{A} - \frac{M^{2}}{2 p}}\,, \\
    \mathcal{A}_{2} =& \frac{M^{2} m^{2}_{\mu s}}{4 p^{2}}\frac{\left(1 - \frac{2 p}{M^{2}}(\tilde{n}\lambda_{2} - \tilde{n}\lambda_{3})\right)^{2}}{V_{NC} + \frac{1}{2}V_{CC} + \frac{1}{2}(\tilde{n}\lambda_{1} + \tilde{n}\lambda_{2})  - \tilde{n}\lambda_{3} + \frac{1}{2}\mathcal{A} - \frac{M^{2}}{2 p}}\,.
\end{align}
The effective mass squared differences are thus
\begin{align}
    \Delta \tilde{m}^{2}_{21} =& 2 p \mathcal{A} + 2p (\mathcal{A}_{2} - \mathcal{A}_{1}) + (m^{2}_{\mu s} - m^{2}_{e s})\left(1 - \frac{2 p}{M^{2}}(\tilde{n}\lambda_{2} - \tilde{n}\lambda_{1})\right)\,, \nonumber \\
    \Delta \tilde{m}^{2}_{31} =&  M^{2} + 2 p \biggl[ \tilde{n}\lambda_{3} - V_{NC} - \frac{1}{2}V_{CC} - \frac{1}{2}(\tilde{n}\lambda_{2}+\tilde{n}\lambda_{1}) + \frac{1}{2}\mathcal{A} - 2 \mathcal{A}_{1} - \mathcal{A}_{2} \biggr] + m^{2}_{\mu s} \nonumber \\
    & + 2 p \left( \frac{m^{2}}{M^{2}} + \frac{m^{2}_{e s}}{M^{2}} \right) (\tilde{n}\lambda_{2} - \tilde{n}\lambda_{3})\,.
    \label{eq:msd_align_2p1}
\end{align}
The effective mass squared differences in Eq. \eqref{eq:msd_align_2p1} in matter are manifestly dependent only on the differences of the torsional coupling constants. Also, we see again that at the high momentum regime, the effective mass squared difference is dominated by the matter effects --- Standard Model and Spin-Torsion. At the same time the effective mass squared differences in matter do not depend on the Majorana phases. {This is consistent with the wisdom that we gathered in the $1+1$ scenario -- the effective mass squared difference is independent of the Majorana phases if the torsion fields are aligned with the mass fields.}

\subsection{Torsion-mass mixing: \texorpdfstring{$\phi_{12} \ne 0$}{phi12 neq 0}}
\label{subsec:phi12_2p1}

In Sec.~\ref{subsec:aligned_2p1} we studied the dispersion relations in matter when the torsion-mass mixing matrix $G = \mathbb{I}_3$, i.e.,  $\phi_{12} = \phi_{13} = \phi_{23} = 0$. In this section, we will consider what happens when the torsion basis is not perfectly aligned with the mass basis. As a simple case, we take $\phi_{12} \ne 0$ and $\phi_{13} = \phi_{23} = 0$. As before we will calculate the energy eigenvalues. The Hamiltonian in this situation is calculated in Appendix~\ref{app:2p1_phi12}. We collected all the terms upto $\mathcal{O}((m/M)^{0})$ in Eq.~\eqref{eq:leading_order_Hprime_phi12_2p1}, from which we can see that the energy eigenvalues correct up to $\mathcal{O}((m/M)^{0})$ are
\begin{align}
    E^{\prime}_{1} =& p + V_{NC} + \frac{1}{2}V_{CC} + \frac{1}{2}(\tilde{n}\lambda_{1} + \tilde{n}\lambda_{2}) - \frac{1}{2}\mathcal{B} \nonumber \\
    E^{\prime}_{2} =& p + V_{NC} + \frac{1}{2}V_{CC} + \frac{1}{2}(\tilde{n}\lambda_{1} + \tilde{n}\lambda_{2}) + \frac{1}{2}\mathcal{B} \nonumber \\
    E^{\prime}_{3} =& p + \frac{M^{2}}{2 p} + \tilde{n}\lambda_{3}\, \label{eq:leading_order_eigenval_phi12_2p1},
\end{align}
where we have defined
\begin{align}
    \mathcal{B} =& \left[ \left(V_{CC} + (\cos (2\phi_{12})(m^{2}_{\mu s}-m^{2}_{e s}) - 2\cos(\alpha - \beta - \phi)\sin (2\phi_{12})m_{e s}m_{\mu s})\frac{(\tilde{n}\lambda_{1}-\tilde{n}\lambda_{2})}{m^{2}}\right)^{2} \right. \nonumber\\
    & \left. + \bigg\lvert \left( 2\cos (2\phi_{12})m_{e s}m_{\mu s} + \sin(2\phi_{12})(e^{i(\alpha - \beta - \phi)}m^{2}_{\mu s} - e^{-i(\alpha - \beta - \phi)}m^{2}_{e s} ) \right) \frac{(\tilde{n}\lambda_{1} - \tilde{n}\lambda_{2})}{m^{2}} \bigg\rvert^{2} \right]^{1/2}\,.
\end{align}
The effective mass squared differences in matter are
\begin{align}
    \Delta \tilde{m}^{2}_{21} =& 2p(E^{\prime}_{2}-E^{\prime}_{1}) = 2 p \mathcal{B} \nonumber \\
    \Delta \tilde{m}^{2}_{31} =& 2p(E^{\prime}_{3} - E^{\prime}_{1}) = M^{2} + 2 p \left[ - V_{NC} - \frac{V_{CC}}{2} + \tilde{n}\lambda_{3} - \frac{\tilde{n}\lambda_{1} + \tilde{n}\lambda_{2}}{2} + \frac{1}{2}\mathcal{B} \right].\label{eq:msd_leading_order_phi12_2p1}
\end{align}
Eq.~\eqref{eq:msd_leading_order_phi12_2p1} gives the effective mass squared difference in the presence of matter. Even without any correction from the $m/M$ terms, we already see that the effective mass squared differences in matter depend on the Majorana phases $\alpha, \beta$ and the phase $\phi$. We can see that $\Delta \tilde{m}^{2}_{(2,3)1}$ are dependent on the Majorana phases and $\phi$ only when {torsion fields and definite mass fields do not align.} This is similar to the behavior of effective mass squared difference in the $1+1$ scenario in Eq.~\eqref{eq:msd_1p1}.
Furthermore, $\Delta \tilde{m}^{2}_{31}$ is not invariant under the transformation $V_{CC} \to -V_{CC}, V_{NC} \to -V_{NC}, \lambda_{A} \to - \lambda_{A}, \alpha \to -\alpha, \beta \to -\beta, \phi \to -\phi$. This is a signature of CP violation. Thus, we see that CP symmetry is broken even when $m/M$ is neglected, i.e. without active sterile mixing, when the torsional four-fermion interaction is taken into account.

We can find the second-order corrections to the energy eigenvalues from the Hamiltonian entries from Eq.~\eqref{eq:second_order_Hprime_phi12_2p1}. The eigenvalues including the second-order corrections are
\begin{align}
    E^{\prime}_{1} =& p + V_{NC} + \frac{1}{2}V_{CC} + \frac{1}{2}(\tilde{n}\lambda_{1} + \tilde{n}\lambda_{2}) - \frac{1}{2}\mathcal{B} + \mathcal{B}_{1} \nonumber \\
    & + \frac{m^{2}_{e s}}{2 p} - \frac{m^{2}_{e s}}{M^{2}}(\sin^{2}\phi_{12}\tilde{n}\lambda_{1} + \cos^{2}\phi_{12}\tilde{n}\lambda_{2} - \tilde{n}\lambda_{3}) + \frac{\cos(\alpha - \beta -\phi)\sin(2\phi_{12})m_{e s}m_{\mu s}}{2 M^{2}}(\tilde{n}\lambda_{1} - \tilde{n}\lambda_{2})\,, \nonumber \\
    E^{\prime}_{2} =& p + V_{NC} + \frac{1}{2}V_{CC} + \frac{1}{2}(\tilde{n}\lambda_{1} + \tilde{n}\lambda_{2}) + \frac{1}{2}\mathcal{B} + \mathcal{B}_{2} \nonumber \\
    & + \frac{m^{2}_{\mu s}}{2 p} - \frac{m^{2}_{\mu s}}{M^{2}}(\sin^{2}\phi_{12}\tilde{n}\lambda_{1} + \cos^{2}\phi_{12}\tilde{n}\lambda_{2} - \tilde{n}\lambda_{3}) - \frac{\cos(\alpha - \beta - \phi)\sin(2\phi_{12})m_{e s}m_{\mu s}}{2 M^{2}}(\tilde{n}\lambda_{1} - \tilde{n}\lambda_{2})\,, \nonumber \\
    E^{\prime}_{3} =& \frac{M^{2}}{2 p} + \tilde{n}\lambda_{3} - \mathcal{B}_{1} - \mathcal{B}_{2} + \frac{m^{2}}{2 p} + \frac{m^{2}}{M^{2}}(\sin^{2}\phi_{12} \tilde{n}\lambda_{1} + \cos^{2}\phi_{12}\tilde{n}\lambda_{2} - \tilde{n}\lambda_{3})\, \label{eq:second_order_eigenval_phi12_2p1}.
\end{align}
Here we have written 
\begin{align}
    \mathcal{B}_{1} =& \frac{1}{4 p^{2} M^{2}}\frac{\left|M^{2}m_{es} - 2 p m_{e s}(\sin^{2}\phi_{12}\tilde{n}\lambda_{1}+\cos^{2}\phi_{12}\tilde{n}\lambda_{2}-\tilde{n}\lambda_{3}) + e^{i(\alpha - \beta - \phi)}p \sin (2 \phi_{12}) m_{\mu s}(\tilde{n}\lambda_{1}-\tilde{n}\lambda_{2})\right|^{2}}{V_{NC} + \frac{1}{2}V_{CC} + \frac{1}{2}(\tilde{n}\lambda_{1}+\tilde{n}\lambda_{2}) -\tilde{n}\lambda_{3} - \frac{1}{2}\mathcal{B} - \frac{M^{2}}{2 p}}\,, \\
    \mathcal{B}_{2} =& \frac{1}{4 p^{2} M^{2}}\frac{\left| M^{2}m_{\mu s} - 2 p m_{\mu s}(\sin^{2}\phi_{12}\tilde{n}\lambda_{1} + \cos^{2}\phi_{12}\tilde{n}\lambda_{2} - \tilde{n}\lambda_{3}) - e^{i(\alpha - \beta - \phi)}p \sin(2\phi_{12})m_{e s}(\tilde{n}\lambda_{1} - \tilde{n}\lambda_{2}) \right|^{2}}{V_{NC} + \frac{1}{2}V_{CC} + \frac{1}{2}(\tilde{n}\lambda_{1}+\tilde{n}\lambda_{2}) -\tilde{n}\lambda_{3} + \frac{1}{2}\mathcal{B} - \frac{M^{2}}{2 p}}\,.
\end{align}
The effective mass squared differences in matter are thus
\begin{align}
    \Delta \tilde{m}^{2}_{21} =& 2p(E^{\prime}_{2} - E^{\prime}_{1}) \notag \\ 
    =& 2 p \mathcal{B} + 2 p (\mathcal{B}_{1} - \mathcal{B}_{2}) + (m^{2}_{\mu s} - m^{2}_{e s})\left( 1 - \frac{2 p}{M^{2}}(\sin^{2}\phi_{12}\tilde{n}\lambda_{1} + \cos^{2}\phi_{12}\tilde{n}\lambda_{2} - \tilde{n}\lambda_{3})\right) \nonumber \\
    & - \frac{2 p \cos (\alpha - \beta - \phi) \sin (2\phi_{12}) m_{e s}m_{\mu s}}{M^{2}}(\tilde{n}\lambda_{1} - \tilde{n}\lambda_{2})\,, \nonumber \\
    \Delta \tilde{m}^{2}_{31} =& 2p(E^{\prime}_{3} - E^{\prime}_{1}) \notag \\
    =& M^{2} + m^{2}_{\mu s} + p \mathcal{B} - 2p (2 \mathcal{B}_{1} + \mathcal{B}_{2}) + 2 p \left( \frac{m^{2}}{M^{2}} + \frac{m^{2}_{e s}}{M^{2}} \right) (\sin^{2}\phi_{12}\tilde{n}\lambda_{1} + \cos^{2}\phi_{12}\tilde{n}\lambda_{2} - \tilde{n}\lambda_{3}) \nonumber \\
    & - \frac{p \cos(\alpha - \beta - \phi) \sin (2\phi_{12}) m_{e s}m_{\mu s}}{M^{2}}(\tilde{n}\lambda_{1} - \tilde{n}\lambda_{2}) + 2p \left( -V_{NC} - \frac{V_{CC}}{2} + \tilde{n}\lambda_{3} - \frac{1}{2}(\tilde{n}\lambda_{1} + \tilde{n}\lambda_{2}) \right)\,.
    \label{eq:msd_phi12_2p1}
\end{align}
As we can see, the corrections in the mass squared differences are second order in $m/M$.
The second order contributions $\mathcal{B}_{1}, \mathcal{B}_{2}$ are not invariant under the transformation --- $V_{CC} \to - V_{CC}, V_{NC} \to -V_{NC}, \lambda_{A} \to -\lambda_{A}, \phi \to -\phi, \alpha \to -\alpha, \beta \to -\beta$, but $\mathcal{B}$ is invariant under this transformation. 
This shows that the second-order corrections also contribute to CP violation. 

\subsection{Active-sterile mixing by torsion: \texorpdfstring{$\phi_{13} \ne 0$}{phi13 neq 0}}
\label{subsec:phi13_2p1}

Above we considered the case when torsion mixes the active neutrino flavor fields and found that this caused CP violation. We will now consider the case when torsion mixes the sterile neutrino flavor with an active neutrino flavor. The Hamiltonian elements are explicitly calculated in Appendix~\ref{app:2p1_phi13}. Above, we evaluated the energy eigenvalues for two cases. In both of these, we defined a dimensionless Hamiltonian $H^{\prime} = (2pH)/M^{2}$ and collected all the terms $\mathcal{O}((m/M)^{0})$ in $(H^{\prime})^{(0)}$. From Eqs.~\eqref{eq:leading_order_Hprime_aligned_2p1} and \eqref{eq:leading_order_Hprime_phi12_2p1} we see that $(H^{\prime})^{(0)}_{e\tau} = 0$ in both these cases. As a result, the characteristic equation can be decomposed into a quadratic and a linear equation, which can be solved easily. This simplification does not occur when $\phi_{13} \neq 0$ -- the corresponding $(H^{\prime})^{(0)}$ is written in Eq.~\eqref{eq:leading_order_Hprime_phi13_2p1}. The leading order energy eigenvalues are the eigenvalues of the $(H^{\prime})^{(0)}$ and are the roots of the characteristic equation 
\begin{align}
    x^{3} - A x^{2} + B x - C =& 0\,, \label{eq:phi13_char_eqn}
\end{align}
with 
\begin{equation}
    A = {\rm Tr}[(H^{\prime})^{(0)}]; \qquad B = \frac{1}{2}\left(({\rm Tr}[(H^{\prime})^{(0)}])^{2} - {\rm Tr}[((H^{\prime})^{(0)})^{2}]\right); \qquad C = {\rm det}[(H^{\prime})^{(0)}]\,.
\end{equation}
Cardano's method is used to solve the cubic equation, giving the energy eigenvalues
\begin{align}
    E^{\prime}_{j} =& \frac{M^{2}}{2 p}\left[\frac{A}{3} + r \cos \left( \frac{1}{3} \arccos \left( - \frac{4 q}{r^3} \right) - \frac{2 \pi j}{3} \right) \right]\,,
\end{align}
with
\begin{equation}
    p = B - \frac{A^{2}}{3}, \qquad q = C + \frac{A B}{3} - \frac{2 A^{3}}{27} \qquad  r = \sqrt{- \frac{4 p}{3}}\,.
\end{equation}
These solutions agree with the solutions given in~\cite{Zaglauer:1988gz, Barger:1980tf, Naumov:1991ju, Kimura:2002hb, Fong:2022oim} if all the Spin-Torsion coupling constants are set to zero. Now we can find out the effective mass squared difference from the energy eigenvalues.

\section{Signatures of CP and T violation}
\label{sec:cp_2p1}

After the success of Long Baseline Experiments, measuring CP phase has become one of the top priorities of the neutrino physics community \cite{Hyper-Kamiokande:2018ofw, DUNE:2020lwj}. In Sec.~\ref{sec:2p1}, we have discussed signatures of CP violation by comparing the effective mass squared differences in matter for neutrinos and antineutrinos. This is not a direct test of CP violation, which would compare neutrino oscillation in matter and antineutrino oscillation in antimatter~\cite{Bitter:2024eka}. The absence of an experimental setup where this can be done prevents us from directly testing CP violation in neutrino oscillations. However, tests of T violation give a direct measure of intrinsic CP violation. The study of T violation in neutrino oscillations is not new~\cite{Cabibbo:1977nk, Kuo:1987km, Krastev:1988yu, Toshev:1989vz, Arafune:1996bt, Harrison:1999df, Akhmedov:2001kd, Minakata:2003jj, Petcov:2018zka} --- it involves comparing $P_{\nu_{\alpha} \to \nu_{\beta}}$ and $P_{\nu_{\beta} \to \nu_{\alpha}}$ in matter, as opposed to comparing $P_{\nu_{\alpha} \to \nu_{\beta}}$ and $P_{\hat{\nu}_{\alpha} \to \hat{\nu}_{\beta}}$, which is done for testing CP violation in neutrino oscillation experiments.

If the density profile along the path from the source to the detector is symmetric under reflection about the midpoint of the path, then T violation implies genuine CP violation, i.e., T invariance holds if the mixing matrix is real~\cite{Kitano:2024kdv, Bitter:2024eka}. In other words, if there is a T violation in such an experiment, the mixing matrix must be complex. Several recent works have started studying T violation in proposed neutrino factories to precisely measure the complex phases in the flavor-mass mixing matrix~\cite{Schwetz:2021cuj, Bitter:2024eka, Kitano:2024kdv}. These and other previous works have considered the T violation in the presence of SM and some generic NSI~\cite{Schwetz:2021thj}. Let us investigate the possibility of T violation in the presence of the torsional four-fermion interaction.

We consider the flavor evolution of a neutrino described by a Hamiltonian $H$ which includes matter effects. Let us consider a unitary matrix $W$ such that $W^{\dagger}H W$ is a diagonal matrix with diagonal elements $E^{\prime}_{i}$. Now the oscillation probability is~\cite{Giunti:2007ry}
\begin{align}
    P_{\nu_{e} \to \nu_{\mu}} = - 4 \sum_{i<j} \Real [W_{e i} W^{*}_{\mu i} W^{*}_{e j} W_{\mu j}] \sin^{2}\left(\frac{(E^{\prime}_{i}-E^{\prime}_{j})L}{2}\right) + 2 \sum_{i<j} \Imag [W_{e i} W^{*}_{\mu i} W^{*}_{e j} W_{\mu j}] \sin ((E^{\prime}_{i}-E^{\prime}_{j})L)\,.
\end{align}
We see that $\Imag [W_{e 1} W^{*}_{\mu 1} W^{*}_{e 2} W_{\mu 2}]$ gives the T violation as measured by $P_{\nu_{\alpha} \to \nu_{\beta}} - P_{\nu_{\beta} \to \nu_{\alpha}}$. It can be shown that~\cite{Kimura:2002hb}
\begin{align}
    \Imag [W_{e 1} W^{*}_{\mu 1} W^{*}_{e 2} W_{\mu 2}] = \frac{\Imag (H_{e \mu} H_{\mu s} H_{s e})}{(E^{\prime}_{1} - E^{\prime}_{2})(E^{\prime}_{2} - E^{\prime}_{3})(E^{\prime}_{3} - E^{\prime}_{1})} = \tilde{J} \label{eq:jarlskog_in_matter}
,\end{align}
the well-known Jarlskog invariant~\cite{Jarlskog:1985ht} which is a measure of T violation. The tilde explicitly reminds us that the $J$ is calculated in matter. 
%
{Evaluating $\tilde{J}$ using the definition of Eq.~\eqref{eq:jarlskog_in_matter} with the Hamiltonian in Eq.~\eqref{eq:H_aligned_2p1} we find}
\begin{align}
    \tilde{J} =& 0\,.
\end{align}
In other words, when the torsion fields are aligned with the definite mass fields, the effective Jarlskog invariant in matter is zero, and hence T-violation is absent.

Let us consider when they are not aligned, say $\phi_{12} \neq 0$. For this we can calculate $\tilde{J}$ using the Hamiltonian in Eq.~\eqref{eq:H_phi12_2p1} to get
\begin{align}
    \tilde{J} \propto& - \frac{\sin(\alpha - \beta - \phi) \sin (2\phi_{12}) m_{e s} m_{\mu s}}{8 M^{2} p^{2}} (\tilde{n}\lambda_{1} - \tilde{n}\lambda_{2})(M^{2} - 2 p (\tilde{n}\lambda_{1} - \tilde{n}\lambda_{3}))(M^{2} - 2 p (\tilde{n}\lambda_{2} - \tilde{n}\lambda_{3}))\,. \label{eq:jarlskog_phi12_2p1}
\end{align}
We see that Jarlskog invariant can be non-zero in this situation. Indeed, we find T-violation whenever the torsion fields are not aligned with the definite mass fields. Evaluating $\tilde{J}$ when only $\phi_{13} \neq 0$ using the Hamiltonian in Eq.~\eqref{eq:H_phi13_2p1} gives 
\begin{align}
    \tilde{J} \propto & - \frac{\cos (\alpha - \phi) \sin (2\phi_{13}) m_{e s}m_{\mu s}}{8 m M p}(\tilde{n}\lambda_{1} - \tilde{n}\lambda_{3}) (2M^{2}(\cos^{2}\phi_{13}\tilde{n}\lambda_{1} - \tilde{n}\lambda_{2} + \sin^{2}\phi_{13}\tilde{n}\lambda_{3}) + 4 p (\tilde{n}\lambda_{2}-\tilde{n}\lambda_{3})(\tilde{n}\lambda_{2}-\tilde{n}\lambda_{1}))\,.
    \label{eq:jarlskog_phi13_2p1}
\end{align}
This again shows that the T violation appears in this scenario essentially due to the non-alignment of torsion fields with the mass fields. We also see that $\tilde{J}$ vanishes in the large $M$ limit. This makes sense physically, as in this limit the sterile neutrino decouples and the oscillation reduces to an effective 2-neutrino oscillation, so that there is no T violation. It can also be seen that $m_{e s}, m_{\mu s}$ have to be both non-zero for non-zero T violation. 


\section{Conclusion}
\label{sec:conclusion}

In this paper we have considered neutrinos with masses generated via the Type I SeeSaw mechanism.
This mechanism enlarges the fermion sector by introducing sterile neutrinos. These sterile neutrinos do not take part in weak interactions, but of course  they interact with the spacetime geometry like any other fermion. The effective four fermion interaction which comes from  torsion also involves sterile neutrinos along with active neutrinos and other fermions, and torsion introduces yet another mixing matrix among flavors. The presence of mixing between the sterile neutrinos and the active neutrinos imposes corrections in the active-active mixing. We have shown this by calculating the effective mass squared differences due to the sterile active mixing up to the second order of the light to heavy neutrino mass ratios. The discussion is carried out in the 1 active neutrino + 1 sterile neutrino and the 2 active neutrinos + 1 sterile neutrino scenario.

We have calculated the expression of the effective mass squared difference in matter in the 1+1 scenario and in two special cases in the 2+1 scenario. The effective mass squared differences explicitly show matter-induced CP violation. We also find that the Majorana phases enter the CP violation only if the torsion fields and mass fields are not aligned. The Dirac mass terms introduce phases in the mass matrix. In the 1+1 scenario, this phase can be absorbed in field redefinitions. However, in the 2+1 scenario, one combination of the two phases cannot be removed. When the torsion fields are aligned with the mass basis, this extra symmetry makes the effective mass squared difference independent of the Dirac phases. When the torsion fields are not aligned with the mass fields, the effective mass squared differences depend on the Dirac phases.

We have also analyzed CP violation or more precisely, T violation, due to the Spin-Torsion coupling in neutrino interactions in the 2+1 scenario. The merits of T violation should be emphasized again. The neutrino mass squared differences are the same for the neutrinos and antineutrinos. However, this is not the case in the presence of matter. Weak interactions introduce CP violation --- the propagation of a neutrino through matter is different from the propagation of an antineutrino through antimatter. CPT symmetry recasts the question into the difference between conversion of $\nu_{\alpha} \to \nu_{\beta}$ and $\nu_{\beta} \to \nu_{\alpha}$. Our inability to construct an experiment through antimatter makes it hard to test CP violation directly in this manner. The conversion process gives an experimentally feasible test of CP violation. We calculate the effective Jarlskog invariant in matter and find the interesting result that matter-induced T violation is present only if there is a non-trivial mixing between torsion fields $\psi_{A}$ and definite mass fields $\psi_{i}$.

In this paper we have worked with torsion fields which are Majorana in nature. We note that for such states, the matter potential induced by this Spin-Torsion interaction contains the contribution of of both $\lambda_{A}^{L}$ and $\lambda_{A}^{R}$ as in Eq.~\eqref{eq:spin-torsion_hamiltonian_majorana}. However, if the torsion states are Dirac in nature, then the contributions of $\lambda_{A}^{L}$ will be suppressed in the $h=+1$ sector, and $\lambda_{A}^{R}$ contributions will be suppressed in the $h = -1$ sector. This is seen from the expressions in Eq.~\eqref{eq:spin-torsion_hamiltonian_dirac}. Also, the matter potentials for Dirac-type torsion states for antineutrinos will be of opposite sign of the neutrinos. For Majorana-type torsion states, there will be no sign change between the neutrino and antineutrino sectors. In the antineutrino sector, the positive helicity sector will be the dominating component. Hence, the Dirac or Majorana nature of the torsion fields will be reflected in the CP violation measures or T violation measures.

Future generation neutrino oscillation experiments will be searching for the existence of sterile neutrinos. If the presence of a sterile neutrino is confirmed, then the finite active-sterile mixing is to be taken into account. Then one has to take into account three active neutrinos with realistic mass models \cite{King:2015dvf, deMedeirosVarzielas:2023ujt, Costa:2023bxw, deMedeirosVarzielas:2022fbw, Zhao:2021tgi}. Such analyses at the probability and event levels will be carried out elsewhere in the future. This work does not take quark mixing into consideration. In the SM the mixing of quarks is governed by the Cabibo-Kobayashi-Maskawa (CKM) matrix and the torsional interaction will also introduce an independent mixing between different generations of quarks. The implications of such mixing remain to be investigated, in particular in the context of nucleon-neutrino interactions.

We have only investigated how Majorana neutrinos interact with torsion in terrestrial environments. How such interactions affect solar neutrinos or neutrinos from supernovae remain to be seen. This work highlights that the spacetime torsion affects CP violation and T violation in neutrino processes. Since future experiments aim to measure CP violation more closely, it is important to include the effects of the torsional interaction in the interpretation of experimental data.
%

\appendix

\section{\texorpdfstring{Derivation of Eqs.~\eqref{eq:spin-torsion_hamiltonian_dirac} and \eqref{eq:spin-torsion_hamiltonian_majorana}}{Derivation of the Eqs. Dirac-torsion and Majorana-torsion}}
\label{app:matter_pot}

The torsion fields that appear in Eq.~\eqref{torsion-H} can be either a Dirac fermion or a Majorana fermion. For a Dirac fermion, its mode expansion is given by
\begin{align}
    \nu_{A}(x) = \int \frac{d^{3}p}{(2\pi)^{3/2}\sqrt{2E_{p}}}\sum_{h = \pm 1} (a^{(h)}_{A}(p)u^{(h)}_{A}(p)e^{-i p\cdot x} + b^{(h)}_{A}{}^{\dagger}(p)v^{(h)}_{A}(p)e^{i p \cdot x})
.\end{align}
Here, $a^{(h)}_{A}(p),~b^{(h)}_{A}(p)$ are the annihilation operators corresponding to the distinct neutrinos and antineutrinos of type $A$ with helicity $h$ and momentum $p$. For a Dirac field, the neutrinos and antineutrinos are distinct. However, for a Majorana field, the mode expansion is given by
\begin{align}
    \nu_{A}(x) = \int \frac{d^{3}p}{(2\pi)^{3/2}\sqrt{2E_{p}}}\sum_{h = \pm 1} (a^{(h)}_{A}(p)u^{(h)}_{A}(p)e^{-i p\cdot x} + a^{(h)}_{A}{}^{\dagger}(p)v^{(h)}_{A}(p)e^{i p \cdot x})
.\end{align}
In the mode expansion of a Majorana field, there can be in general a phase factor, say in front of the creation operator, which will never appear in a physical amplitude \cite{Pal:2014xrq}. Hence, we drop that phase from the mode expansion. Here, $E_{p}$ is the on-shell energy and is given by $E_{p} = \sqrt{|\vec{p}|^{2} + m_{A}^{2}}$, where $m_{A}$ is the mass of the neutrino type $A$ if the mixing is ignored. 

The elements of the Hamiltonian calculated from Eq.~\eqref{torsion-H} are given by
\begin{align}
    (H_{\text{eff}}^{\text{ST}})^{h h^{\prime}}_{AA} = \bra{\nu^{(h)}_{A}} \int d^{3}x \Ha_{\text{eff}}^{ST} \ket{\nu^{(h^{\prime})}_{A}}
,\end{align}
where 
\begin{align}
    \ket{\nu^{(h)}_{A}} = \sqrt{\frac{(2\pi)^{
    3}}{V}}a^{(h)}_{A}{}^{\dagger}(p)\ket{0}
.\end{align}
For Dirac-type torsion fields
\begin{align}
    (H_{\text{eff}}^{\text{ST}})^{h h^{\prime}}_{AA} = \bar{u}^{h}_{A}(p)\gamma^{0}(1-\gamma^{5})u^{(h^{\prime})}_{A}(p)\delta_{h h^{\prime}}
,\end{align}
while for Majorana-type torsion fields
\begin{align}
    (H_{\text{eff}}^{\text{ST}})^{h h^{\prime}}_{AA} = (\bar{u}^{h}_{A}(p)\gamma^{0}(1-\gamma^{5})u^{(h^{\prime})}_{A}(p) - \bar{v}^{h}_{A}(p)\gamma^{0}(1-\gamma^{5})v^{(h^{\prime})}_{A}(p))\delta_{h h^{\prime}}
.\end{align}
We will now make use of the relations
\begin{align}
    u^{(h)}_{A}(p) \bar{u}^{(h)}_{A}(p) = (\slashed{p}+m_{A})\left(\frac{1 + \gamma^{5}\slashed{s}_{h}}{2}\right), \qquad v^{(h)}_{A}(p) \bar{v}^{(h)}_{A}(p) = (\slashed{p}-m_{A})\left(\frac{1 + \gamma^{5}\slashed{s}_{h}}{2}\right)
,\end{align}
where
\begin{align}
    s^{\mu}_{h}=h\left(\frac{|\vec{p}|}{m},\frac{E}{m}\frac{\vec{p}}{|\vec{p}|}\right)
.\end{align}
Using this and some standard trace relations involving Dirac gamma matrices, we get
\begin{align}
    (H)^{h h^{\prime}}_{AA} = \tilde{n}\left(\frac{E_{p} - h |\vec{p}|}{2 E_{p}}\lambda^{L}_{A} + \frac{E_{p} + h |\vec{p}|}{2 E_{p}}\lambda^{R}_{A}\right) \label{eq:actual_spin-torsion_hamiltonian_dirac}
\end{align}
for Dirac-type torsion states, and 
\begin{align}
    (H)^{h h^{\prime}}_{AA} = -\tilde{n}\frac{h |\vec{p}|}{2 E_{p}}(\lambda^{L}_{A} - \lambda^{R}_{A})\delta_{h h^{\prime}} \label{eq:actual_spin-torsion_hamiltonian_majorana}
\end{align}
for Majorana-type torsion states. Eqs.~\eqref{eq:actual_spin-torsion_hamiltonian_dirac} and \eqref{eq:actual_spin-torsion_hamiltonian_majorana} approximate to Eqs.~\eqref{eq:spin-torsion_hamiltonian_dirac} and \eqref{eq:spin-torsion_hamiltonian_majorana} respectively in the ultrarelativistic limit. 

\section{Decoupling the two helicities in the neutrino Hamiltonian}
\label{sec:helicity}

Following \cite{Smirnov:2025wax, Akhmedov:2025enm}, we now show how the two helicities decouple for ultrarelativistic neutrinos. The Schr\"odinger equation from Eq.~\eqref{eq:schro_full} when the torsion states are Dirac is given by
%
\begin{align}
i\frac{d}{dt}\begin{pmatrix}c^{(-)}\\ c^{(+)}\end{pmatrix} =& \left[\begin{pmatrix}\tilde{U}E\tilde{U}^{\dagger} & 0\\ 0 & \tilde{U}E\tilde{U}^{\dagger}\end{pmatrix}
+ \begin{pmatrix}V_{SM} & 0\\ 0 & V_{SM}\mathcal{O}(\frac{m^{2}}{E^{2}})\end{pmatrix} \right. \nonumber \\
& \left.+ \tilde{n}\begin{pmatrix}(\tilde{U}G)(\lambda^{L} + \lambda^{R}\mathcal{O}(\frac{m^{2}}{E^{2}}))(\tilde{U}G)^{\dagger} & 0\\ 
 0 & (\tilde{U}G)(\lambda^{R} + \lambda^{L}\mathcal{O}(\frac{m^{2}}{E^{2}}))(\tilde{U}G)^{\dagger}\end{pmatrix} \right] \begin{pmatrix}
 c^{(-)}\\ c^{(+)}\end{pmatrix} \nonumber \\
i\frac{d}{dt}\begin{pmatrix}c^{(-)}\\ c^{(+)}\end{pmatrix} =& H  \begin{pmatrix} c^{(-)}\\ c^{(+)}\end{pmatrix} \label{eq:flavor_eqn_full}
.\end{align}
Here the $c^{(-)}$ is a $N_{a} + N$ dimensional column vector of $c^{(-)}_{\alpha}$s, a similar definition is applicable for $c_{\alpha}^{(+)}$ as well.
%
Suppose the $H$ of Eq.~\eqref{eq:flavor_eqn_full} can be diagonalized by a matrix 
\begin{align}
 \begin{pmatrix} W^{(-)} & 0\\ 0 & W^{(+)}\end{pmatrix}
.\end{align}
As there are no mixing between two helicity sectors, we can write the diagonalizing
matrix in block diagonal form. The 
formal solution to the above equation in ultrarelativistic limit is 
\begin{align}
\begin{pmatrix}c^{(-)}(L)\\ c^{(+)}(L)\end{pmatrix} = \begin{pmatrix} W^{(-)} & 0\\ 0 & W^{(+)}\end{pmatrix}
\exp\left[-i \begin{pmatrix} \hat{H}^{(-)} & 0\\ 0 & \hat{H}^{(+)} \end{pmatrix}L \right]
\begin{pmatrix} (W^{(-)})^{\dagger} & 0\\ 0 & (W^{(+)})^{\dagger}\end{pmatrix}\begin{pmatrix}c^{(-)}(0)\\ c^{(+)}(0)\end{pmatrix}
.\end{align}
Here, $L$ is the distance traversed by neutrinos. $\hat{H}^{(h)}$ is the diagonalized Hamiltonian for the helicity sector $h$. Let us consider an experiment where the Weak interaction produces a neutrino in a state 
$\ket{\psi_{P}} = c^{(-)}_{\alpha}\ket{\nu^{(-)}_{\alpha}} + c^{(+)}_{\alpha}\ket{\nu^{(+)}_{\alpha}}$. Due to the presence of left-chiral projection operator $c^{(+)}$ will be small for ultrarelativistic neutrinos. If the detector wants to detect a neutrino in a state $\ket{\psi_{D}} = d^{(-)}_{\beta}\ket{\nu^{(-)}_{\beta}} + d^{(+)}_{\beta}\ket{\nu^{(+)}_{\beta}}$ then the measurement operator is 
$\ket{\psi_{D}}\bra{\psi_{D}}$. Here, $\beta$ is the flavor of the detected state. The probability of a neutrino produced in state $\ket{\psi_{p}}$ to be detected in state $\ket{\psi_{D}}$ is
\begin{align}
P_{\nu_{\alpha \Le} \to \nu_{\beta \Ri}} = |d^{(-)}_{\beta}[W^{(-)}e^{-i\hat{H}^{(-)}L}(W^{(-)})^{\dagger}]_{\beta \alpha}c^{(-)}_{\alpha} 
+ d^{(+)}_{\beta}[W^{(+)}e^{-i\hat{H}^{(+)}L}(W^{(+)})^{\dagger}]_{\beta \alpha}c^{(+)}_{\alpha}|^{2}
.\end{align}
If the detection operator corresponds to a right-chiral neutrino then $d^{(-)}$ will be much smaller than $d^{(+)}$. So, in the expression of probability both the first and second term are suppressed in the ultrarelativistic limit. The first term is suppressed by $d^{(-)}_{\beta}$ and the second term is suppressed by $c^{(+)}_{\alpha}$. So, we can ignore the chirality flip probability compared to active neutrino transitions.

Similarly, for antineutrinos the Schr\"{o}dinger equation is given by
\begin{align}
i\frac{d}{dt}\begin{pmatrix}\bar{c}^{(-)}\\ \bar{c}^{(+)}\end{pmatrix} =& \left[\begin{pmatrix}\tilde{U}E\tilde{U}^{\dagger} & 0\\ 0 & \tilde{U}E\tilde{U}^{\dagger}\end{pmatrix}
+ \begin{pmatrix}-V_{SM}\mathcal{O}(\frac{m^{2}}{E^{2}}) & 0\\ 0 & -V_{SM}\end{pmatrix} \right. \nonumber \\
& \left.+ \tilde{n}\begin{pmatrix}-(\tilde{U}G)(\lambda^{R} + \lambda^{L}\mathcal{O}(\frac{m^{2}}{E^{2}}))(\tilde{U}G)^{\dagger} & 0\\ 
 0 & -(\tilde{U}G)(\lambda^{L} + \lambda^{R}\mathcal{O}(\frac{m^{2}}{E^{2}}))(\tilde{U}G)^{\dagger} \end{pmatrix} \right] \begin{pmatrix}
 \bar{c}^{(-)}\\ \bar{c}^{(+)}\end{pmatrix} \nonumber \\
i\frac{d}{dt}\begin{pmatrix}\bar{c}^{(-)}\\ \bar{c}^{(+)}\end{pmatrix} =& \bar{H}  \begin{pmatrix}
 \bar{c}^{(-)}\\ \bar{c}^{(+)}\end{pmatrix}
.\end{align}
Now, the prodcution state of antineutrino $\bar{\nu}_{\alpha}$ will have large $\bar{c}^{(+)}_{\alpha}$ but very small 
$\bar{c}^{(-)}_{\alpha}$.
Similarly we can show that owing to the V--A structure of the Weak interaction one of the components in the production and detection state will be suppressed. Hence, the chirality flip probabilities will be suppressed. Similar arguments can be applied when the torsion fields are Majorana.

This means we can study the evolution of two helicities independently as a first approximation in the ultrarelativistic limit.

\section{\texorpdfstring{Calculation of the Hamiltonian for Sec.~\ref{subsec:aligned_2p1}}{Calculation of the Hamiltonian for Sec. aligned-2p1}}
\label{app:2p1_aligned}

In this appendix, we give the details of the calculation of the approximate eigenvalues from the time-independent perturbation theory in the case of Sec.~\ref{subsec:aligned_2p1}. Computing the Hamiltonian in the RHS of Eq.~\eqref{eq:schro_eqn_2p1} for $\phi_{12} = \phi_{13} = \phi_{23} = 0$ and writing
\begin{align}
    H = p + \begin{pmatrix}H_{ee} & H_{e \mu} & H_{e \tau}\\ H^{*}_{e \mu} & H_{\mu \mu} & H_{\mu \tau}\\ H^{*}_{e \tau} & H^{*}_{\mu \tau} & H_{\tau}\end{pmatrix}
,\end{align}
we get
\begin{align}
    H_{e e} =& \frac{m^{2}_{e s}}{2 p} + V_{CC} + V_{NC} + \frac{m^{2}_{\mu s}}{m^{2}}\tilde{n}\lambda_{1} + \frac{m^{2}_{e s}}{m^{2}}(1 - \frac{m^{2}}{M^{2}})\tilde{n}\lambda_{2} + \frac{m^{2}_{e s}}{M^{2}}\tilde{n}\lambda_{3}, \nonumber \\
    H_{e \mu} =& e^{i(\phi_{1}-\phi_{2})} \left( \frac{m_{e s} m_{\mu s}}{2 p} - \frac{m_{e s}m_{\mu s}}{m^{2}}\tilde{n}\lambda_{1} + \frac{m_{e s}m_{\mu s}}{m^{2}}(1 - \frac{m^{2}}{M^{2}})\tilde{n}\lambda_{2} + \frac{m_{e s}m_{\mu s}}{M^{2}}\tilde{n}\lambda_{3} \right), \nonumber \\
    H_{e \tau} =& e^{i\phi_{1}}\left(\frac{M m_{e s}}{2 p} - \frac{m_{e s}}{M}\tilde{n}\lambda_{2} + \frac{m_{e s}}{M}\tilde{n}\lambda_{3} \right), \nonumber \\
    H_{\mu \mu} =& \frac{m^{2}_{\mu s}}{2 p} + V_{NC} + \frac{m^{2}_{e s}}{m^{2}}\tilde{n}\lambda_{1} + \frac{m^{2}_{\mu s}}{m^{2}}(1 - \frac{m^{2}}{M^{2}})\tilde{n}\lambda_{2} + \frac{m^{2}_{\mu s}}{M^{2}}\tilde{n}\lambda_{3}, \nonumber \\
    H_{\mu \tau} =& e^{i\phi_{2}}\left(\frac{M m_{\mu s}}{2 p} - \frac{m_{\mu s}}{M}\tilde{n}\lambda_{2} + \frac{m_{\mu s}}{M}\tilde{n}\lambda_{3}\right), \nonumber \\
    H_{\tau \tau} =& \frac{m^{2}}{2 p} + \frac{M^{2}}{2 p} + \frac{m^{2}}{M^{2}}\tilde{n}\lambda_{2} + \tilde{n}\lambda_{3} - \frac{m^{2}}{M^{2}}\tilde{n}\lambda_{3} \label{eq:H_aligned_2p1}
.\end{align}
In the SeeSaw limit, the small parameters are $m_{e s}/ M, m_{\mu s}/ M, m/ M$. We define a dimensionless Hamiltonian $H^{\prime} = (2pH)/(M^{2})$ and suppress the leading $p$. Now we separate the $\mathcal{O}((m/M)^{0})$ terms of $H^{\prime}$,
\begin{align}
    (H^{\prime})^{(0)}_{e e} =& \frac{2 p}{M^{2}}(V_{CC} + V_{NC}) + \frac{2 p m^{2}_{\mu s}}{m^{2} M^{2}}\tilde{n}\lambda_{1} + \frac{2 p m^{2}_{e s}}{m^{2} M^{2}}\tilde{n}\lambda_{2}, \nonumber \\
    (H^{\prime})^{(0)}_{e \mu} =& -e^{i(\phi_{1}-\phi_{2})}\frac{2 p m_{e s}m_{\mu s}}{m^{2}M^{2}}(\tilde{n}\lambda_{1} - \tilde{n}\lambda_{2}), \nonumber \\
    (H^{\prime})^{(0)}_{e \tau} =& 0, \nonumber \\
    (H^{\prime})^{(0)}_{\mu \mu} =& \frac{2 p}{M^{2}}V_{NC} + \frac{2 p m^{2}_{e s}}{m^{2}M^{2}}\tilde{n}\lambda_{1} + \frac{2 p m^{2}_{\mu s}}{m^{2}M^{2}}\tilde{n}\lambda_{2}, \nonumber \\
    (H^{\prime})^{(0)}_{\mu \tau} =& 0, \nonumber \\
    (H^{\prime})^{(0)}_{\tau \tau} =& 1 + \frac{2 p}{M^{2}}\tilde{n}\lambda_{3}. \label{eq:leading_order_Hprime_aligned_2p1}
\end{align}
The terms of $\mathcal{O}((m/M))$ of $H^{\prime}$ are gathered below
\begin{align}
    (H^{\prime})^{(1)}_{e e} =& 0, \nonumber \\
    (H^{\prime})^{(1)}_{e \mu} =& 0, \nonumber \\
    (H^{\prime})^{(1)}_{e \tau} =& e^{i\phi_{1}}\left(\frac{m_{e s}}{M} - \frac{2 p m_{e s}}{M^{3}}\tilde{n}\lambda_{2} + \frac{2 p m_{e s}}{M^{3}}\tilde{n}\lambda_{3}\right), \nonumber \\
    (H^{\prime})^{(1)}_{\mu \mu} =& 0, \nonumber \\
    (H^{\prime})^{(1)}_{\mu \tau} =& e^{i\phi_{2}}\left(\frac{m_{\mu s}}{M} - \frac{2 p m_{\mu s}}{M^{3}}\tilde{n}\lambda_{2} + \frac{2 p m_{\mu s}}{M^{3}}\tilde{n}\lambda_{3}\right), \nonumber \\
    (H^{\prime})^{(1)}_{\tau \tau} =& 0 \label{eq:first_order_Hprime_aligned_2p1}
.\end{align}
Finally, gathering all the terms of $\mathcal{O}((m/M)^{2})$ in the $H^{\prime}$, we get
\begin{align}
    (H^{\prime})^{(2)}_{e e} =& \frac{m^{2}_{e s}}{M^{2}} - \frac{2 p m^{2}_{e s}}{M^{4}}\tilde{n}\lambda_{2} + \frac{2 p m^{2}_{e s}}{M^{4}}\tilde{n}\lambda_{3}, \nonumber \\
    (H^{\prime})^{(2)}_{e \mu} =& e^{i(\phi_{1}-\phi_{2})}\left(\frac{m_{e s}m_{\mu s}}{M^{2}} - \frac{2 p m_{e s}m_{\mu s}}{M^{4}}\tilde{n}\lambda_{2} + \frac{2 p m_{e s}m_{\mu s}}{M^{4}}\tilde{n}\lambda_{3}\right), \nonumber \\
    (H^{\prime})^{(2)}_{e \tau} =& 0, \nonumber \\
    (H^{\prime})^{(2)}_{\mu \mu} =& \frac{m^{2}_{\mu s}}{M^{2}} - \frac{2 p m^{2}_{\mu s}}{M^{4}}\tilde{n}\lambda_{2} + \frac{2 p m^{2}_{\mu s}}{M^{4}}\tilde{n}\lambda_{3}, \nonumber \\
    (H^{\prime})^{(2)}_{\mu \tau} =& 0, \nonumber \\
    (H^{\prime})^{(2)}_{\tau \tau} =& \frac{m^{2}}{M^{2}} + \frac{2 p m^{2}}{M^{4}}\tilde{n}\lambda_{2} - \frac{2 p m^{2}}{M^{4}}\tilde{n}\lambda_{3} \label{eq:second_order_Hprime_aligned_2p1}
.\end{align}

We use the standard perturbation theory to calculate the approximate energy eigenvalues. The leading order energy eigenvalues are the eigenvalues of $(H^{\prime})^{(0)}$ and is given in Eq.~\eqref{eq:leading_order_eigenval_align_2p1}. There are no first-order corrections to the energy eigenvalues. The second-order corrections to the energy eigenvalues are included in Eq.~\eqref{eq:second_order_eigenval_align_2p1}.

\section{\texorpdfstring{Calculation of the Hamiltonian for Sec.~\ref{subsec:phi12_2p1}}{Calculation of the Hamiltonian for Sec. phi12-2p1}}
\label{app:2p1_phi12}

In this appendix, we will give details of the calculation in Sec.~\ref{subsec:phi12_2p1}. Computing the Hamiltonian in the RHS of Eq.~\eqref{eq:schro_eqn_2p1} for $\phi_{12} \neq 0, \phi_{13} = \phi_{23} = 0$ and writing
\begin{align}
    H = p + \begin{pmatrix}H_{e e} & H_{e \mu} & H_{e \tau}\\ H^{*}_{e \mu} & H_{\mu \mu} & H_{\mu \tau}\\ H^{*}_{e \tau} & H^{*}_{\mu \tau} & H_{\tau \tau}\end{pmatrix}
,\end{align}
we get
\begin{align}
    H_{e e} =& \frac{m^{2}_{e s}}{2 p} + V_{CC} + V_{NC} + \frac{\cos^{2} \phi_{12}m^{2}_{\mu s}}{m^{2}}\tilde{n}\lambda_{1} + \frac{\sin^{2}\phi_{12}m^{2}_{e s}}{m^{2}}\left(1 - \frac{m^{2}}{M^{2}}\right)\tilde{n}\lambda_{1} \notag \\
    & - \frac{\cos (\alpha - \beta - \phi)\sin (2 \phi_{12})m_{e s}m_{\mu s}}{m^{2}}\left(1 - \frac{1}{2}\frac{m^{2}}{M^{2}}\right)\tilde{n}\lambda_{1} + \frac{\sin^{2}\phi_{12}m^{2}_{\mu s}}{m^{2}}\tilde{n}\lambda_{2} + \frac{\cos^{2}\phi_{12}m^{2}_{e s}}{m^{2}}\left(1 - \frac{m^{2}}{M^{2}}\right)\tilde{n}\lambda_{2} \notag \\
    & + \frac{\cos (\alpha - \beta - \phi) \sin(2\phi_{12}) m_{e s}m_{\mu s}}{m^{2}}\left(1 - \frac{1}{2}\frac{m^{2}}{M^{2}}\right)\tilde{n}\lambda_{2} + \frac{m^{2}_{e s}}{M^{2}}\tilde{n}\lambda_{3}, \notag \\
    H_{e \mu} =& \frac{1}{8 m^{2} M^{2} p} e^{i(\phi_{1}-\phi_{2})}\Big(-2p\,e^{-i(\alpha-\beta-\phi_{1}-\phi_{2})}\sin(2\phi_{12})\,m_{e s}^{4}(\tilde{n}\lambda_{1}-\tilde{n}\lambda_{2}) \notag \\
    & +2p\,e^{i(\alpha-\beta-\phi_{1}-\phi_{2})}\sin(2\phi_{12})\,m_{\mu s}^{2}(-2M^{2}+m_{\mu s}^{2})(\tilde{n}\lambda_{1}-\tilde{n}\lambda_{2})+4p\,\sin(2\phi_{12})\,m_{e s}^{2}\big(M^{2}e^{-i(\alpha-\beta-\phi_{1}-\phi_{2})} \notag \\
    & +i\,m_{\mu s}^{2}\sin(\alpha-\beta-\phi_{1}-\phi_{2})\big)(\tilde{n}\lambda_{1}-\tilde{n}\lambda_{2})-8p\,m_{e s}^{3}m_{\mu s}\big(\sin^{2}(\phi_{12})\tilde{n}\lambda_{1}+\cos^{2}(\phi_{12})\tilde{n}\lambda_{2}\big) \notag \\
    & +4 m_{e s}m_{\mu s}\big(-2p(M^{2}\cos(2\phi_{12})+\sin^{2}(\phi_{12})m_{\mu s}^{2})\tilde{n}\lambda_{1}+2p(M^{2}\cos(2\phi_{12})-\cos^{2}(\phi_{12})m_{\mu s}^{2})\tilde{n}\lambda_{2} \notag \\
    & +m^{2}(M^{2}+2p\tilde{n}\lambda_{3})\big)\Big), \notag \\
    H_{e \tau} =& e^{i\phi_{1}}\biggl[ \frac{M m_{e s}}{2 p} - \frac{\sin^{2}\phi_{12}m_{e s}}{M}\tilde{n}\lambda_{1} + e^{i(\alpha - \beta - \phi)} \frac{\sin (2\phi_{12})m_{\mu s}}{2 M}\tilde{n}\lambda_{1} - \frac{\cos^{2}\phi_{12}m_{e s}}{M}\tilde{n}\lambda_{2} \notag \\
    & - e^{i(\alpha - \beta - \phi)}\frac{\sin (2\phi_{12})m_{\mu s}}{2 M}\tilde{n}\lambda_{2} + \frac{m_{e s}}{M}\tilde{n}\lambda_{3} \biggr], \notag \\
    H_{\mu \mu} =& \frac{m^{2}_{\mu s}}{2 p} + V_{NC} + \frac{\cos^{2}\phi_{12}m^{2}_{e s}}{m^{2}}\tilde{n}\lambda_{1} + \frac{\sin^{2}\phi_{12} m^{2}_{\mu s}}{m^{2}}\left(1-\frac{m^{2}}{M^{2}}\right)\tilde{n}\lambda_{1} \notag \\
    & + \frac{\cos (\alpha - \beta - \phi)\sin (2\phi_{12})m_{e s}m_{\mu s}}{m^{2}}\left(1 - \frac{1}{2}\frac{m^{2}}{M^{2}}\right)\tilde{n}\lambda_{1} + \frac{\sin^{2}\phi_{12}m^{2}_{es}}{m^{2}}\tilde{n}\lambda_{2} + \frac{\cos^{2}\phi_{12}m^{2}_{\mu s}}{m^{2}}\left(1-\frac{m^{2}}{M^{2}}\right)\tilde{n}\lambda_{2} \notag \\
    & - \frac{\cos (\alpha - \beta - \phi) \sin (2\phi_{12})m_{e s}m_{\mu s}}{m^{2}}\left(1-\frac{1}{2}\frac{m^{2}}{M^{2}}\right)\tilde{n}\lambda_{2} + \frac{m^{2}_{\mu s}}{M^{2}}\tilde{n}\lambda_{3}, \notag \\
    H_{\mu \tau} =& e^{i\phi_{2}}\biggl[ \frac{M m_{\mu s}}{2 p} - e^{i(\alpha - \beta - \phi)}\frac{\sin (2\phi_{12})m_{e s}}{2 M}\tilde{n}\lambda_{1} - \frac{\sin^{2}\phi_{12}m_{\mu s}}{M}\tilde{n}\lambda_{1} + e^{i(\alpha - \beta -\phi)}\frac{\sin (2\phi_{12})m_{e s}}{2M}\tilde{n}\lambda_{2} \notag \\
    & - \frac{\cos^{2}\phi_{12}m_{\mu s}}{M}\tilde{n}\lambda_{2} + \frac{m_{\mu s}}{M}\tilde{n}\lambda_{3} \biggr], \notag \\
    H_{\tau \tau} =& \frac{M^{2}}{2 p} + \frac{m^{2}}{2 p} + \frac{m^{2}}{2 M^{2}}(1 - \cos(2\phi_{12}))\tilde{n}\lambda_{1} + \frac{m^{2}}{2M^{2}}(1+\cos(2\phi_{12}))\tilde{n}\lambda_{2} + \tilde{n}\lambda_{3} - \frac{m^{2}}{M^{2}}\tilde{n}\lambda_{3} \label{eq:H_phi12_2p1}
.\end{align}
We again define a dimensionless Hamiltonian $H^{\prime}$ such that $H = (M^{2}H^{\prime})/(2p)$ and ignore the leading $p$. Then we split the Hamiltonian entries in order of smallness. Collecting all the terms of $\mathcal{O}((m/M)^{0})$ we get
\begin{align}
    (H^{\prime})^{(0)}_{e e} =&  \frac{2p}{M^{2}}(V_{CC} + V_{NC}) + \frac{2 p\cos^{2} \phi_{12}m^{2}_{\mu s}}{m^{2}M^{2}}\tilde{n}\lambda_{1} + \frac{2 p\sin^{2} \phi_{12}m^{2}_{e s}}{m^{2}M^{2}}\tilde{n}\lambda_{1} + \frac{2 p \sin^{2}\phi_{12}m^{2}_{\mu s}}{m^{2} M^{2}}\tilde{n}\lambda_{2} + \frac{2 p \cos^{2}\phi_{12}m^{2}_{e s}}{m^{2} M^{2}}\tilde{n}\lambda_{2} \nonumber \\
    & - \frac{2 p\cos (\alpha - \beta - \phi)\sin (2 \phi_{12})m_{e s}m_{\mu s}}{m^{2}M^{2}}(\tilde{n}\lambda_{1} - \tilde{n}\lambda_{2}), \nonumber \\
    (H^{\prime})^{(0)}_{e \mu} =& e^{i(\phi_{1}-\phi_{2})}\left( e^{-i(\alpha - \beta - \phi)}\frac{p \sin(2\phi_{12}) m^{2}_{e s}}{m^{2}M^{2}} - e^{i(\alpha - \beta -\phi)}\frac{p \sin(2\phi_{12}) m^{2}_{\mu s}}{m^{2}M^{2}} -\frac{2 p \cos(2\phi_{12}) m_{e s}m_{\mu s}}{m^{2}M^{2}} \right)(\tilde{n}\lambda_{1}-\tilde{n}\lambda_{2}), \nonumber \\
    (H^{\prime})^{(0)}_{e \tau} =& 0, \nonumber \\
    (H^{\prime})^{(0)}_{\mu \mu} =&  \frac{ 2 p V_{NC}}{M^{2}} + \frac{2 p \cos^{2}\phi_{12}m^{2}_{e s}}{m^{2}M^{2}}\tilde{n}\lambda_{1} + \frac{2 p \sin^{2}\phi_{12} m^{2}_{\mu s}}{m^{2} M^{2}}\tilde{n}\lambda_{1} + \frac{2 p \sin^{2}\phi_{12}m^{2}_{es}}{m^{2}M^{2}}\tilde{n}\lambda_{2} + \frac{2 p \cos^{2}\phi_{12}m^{2}_{\mu s}}{m^{2}M^{2}}\tilde{n}\lambda_{2} \nonumber \\
    & + \frac{2 p \cos (\alpha - \beta - \phi)\sin (2\phi_{12})m_{e s}m_{\mu s}}{m^{2}M^{2}}(\tilde{n}\lambda_{1} - \tilde{n}\lambda_{2}), \nonumber \\
    (H^{\prime})^{(0)}_{\mu \tau} =& 0, \nonumber \\
    (H^{\prime})^{(0)}_{\tau \tau} =& 1 + \frac{2 p\tilde{n}\lambda_{3}}{M^{2}}
    \label{eq:leading_order_Hprime_phi12_2p1}
.\end{align}
Collecting all the terms of $\mathcal{O}((m/M))$ of $H^{\prime}$, we get
\begin{align}
    (H^{\prime})^{(1)}_{e e} =& 0, \nonumber \\
    (H^{\prime})^{(1)}_{e \mu} =& 0, \nonumber \\
    (H^{\prime})^{(1)}_{e \tau} =& e^{i\phi_{1}}\biggl[ \frac{m_{e s}}{M} - \frac{2 p \sin^{2}\phi_{12}m_{e s}}{M^{3}}\tilde{n}\lambda_{1} + e^{i(\alpha - \beta - \phi)}\frac{p \sin (2\phi_{12})m_{\mu s}}{M^{3}}\tilde{n}\lambda_{1} - \frac{2 p \cos^{2}\phi_{12} m_{e s}}{M^{3}}\tilde{n}\lambda_{2} \nonumber \\
    & - e^{i(\alpha - \beta - \phi)}\frac{p \sin (2\phi_{12}) m_{\mu s}}{M^{3}}\tilde{n}\lambda_{2} + \frac{2 p m_{e s}}{M^{3}}\tilde{n}\lambda_{3} \biggr], \nonumber \\
    (H^{\prime})^{(1)}_{\mu \mu} =& 0, \nonumber \\
    (H^{\prime})^{(1)}_{\mu \tau} =& e^{i \phi_{2}} \biggl[ \frac{m_{\mu s}}{M} - e^{i(\alpha - \beta - \phi)}\frac{p \sin (2\phi_{12}) m_{e s}}{M^{3}}\tilde{n}\lambda_{1} - \frac{2 p \sin^{2}\phi_{12} m_{\mu s}}{M^{3}}\tilde{n}\lambda_{1} + e^{i(\alpha - \beta - \phi)}\frac{p \sin(2\phi_{12})m_{e s}}{M^{3}}\tilde{n}\lambda_{2}, \nonumber \\
    & - \frac{2 p \cos^{2}\phi_{12}m_{\mu s}}{M^{3}}\tilde{n}\lambda_{2} + \frac{2 p m_{\mu s}}{M^{3}}\tilde{n}\lambda_{3} \biggr], \nonumber \\
    (H^{\prime})^{(1)}_{\tau \tau} =& 0 \label{eq:first_order_Hprime_phi12_2p1}
.\end{align}
All the terms of $\mathcal{O}((m/M)^{2})$ of the $H^{\prime}$ are given by
\begin{align}
    (H^{\prime})^{(2)}_{e e} =& \frac{m^{2}_{e s}}{M^{2}} - \frac{2 p \sin^{2}\phi_{12} m^{2}_{e s}}{M^{4}}\tilde{n}\lambda_{1} - \frac{2 p \cos^{2} \phi_{12} m^{2}_{e s}}{M^{4}}\tilde{n}\lambda_{2} + \frac{p \cos (\alpha - \beta - \phi) \sin(2 \phi_{12}) m_{e s}m_{\mu s}}{M^{4}}(\tilde{n}\lambda_{1} - \tilde{n}\lambda_{2}) \nonumber \\
    & + \frac{2 p m^{2}_{e s}}{M^{4}}\tilde{n}\lambda_{3}, \nonumber \\
    (H^{\prime})^{(2)}_{e \mu} =& e^{i(\phi_{1}-\phi_{2})}\biggl[ - e^{-i(\alpha - \beta - \phi)}\frac{p \sin (2\phi_{12})m^{4}_{e s}}{2 m^{2} M^{4}}(\tilde{n}\lambda_{1} - \tilde{n}\lambda_{2}) + e^{i(\alpha - \beta -\phi)}\frac{p \sin(2\phi_{12})m^{4}_{\mu s}}{m^{2}M^{4}}(\tilde{n}\lambda_{1} - \tilde{n}\lambda_{2}) \nonumber \\
    & + i\sin(\alpha - \beta -\phi) \frac{p \sin(2\phi_{12})m^{2}_{e s}m^{2}_{\mu s}}{m^{2}M^{4}}(\tilde{n}\lambda_{1} - \tilde{n}\lambda_{2}) - \frac{2 p m_{e s} m_{\mu s}}{M^{4}}(\sin^{2}\phi_{12}\tilde{n}\lambda_{1} + \cos^{2}\phi_{12}\tilde{n}\lambda_{2}) \nonumber \\
    & + \frac{m_{e s}m_{\mu s}}{M^{2}} + \frac{2 p m_{e s}m_{\mu s}}{M^{4}}\tilde{n} \lambda_{3} \biggr], \nonumber \\
    (H^{\prime})^{(2)}_{e \tau} =& 0, \nonumber \\
    (H^{\prime})^{(2)}_{\mu\mu} =& \frac{m^{2}_{\mu s}}{M^{2}} - \frac{2 p \sin^{2} \phi_{2} m^{2}_{\mu s}}{M^{4}}\tilde{n}\lambda_{1} - \frac{2 p \cos^{2}\phi_{12} m^{2}_{\mu s}}{M^{4}}\tilde{n}\lambda_{2} - \frac{p \cos (\alpha - \beta - \phi) \sin(2\phi_{12}) m_{e s}m_{\mu s}}{M^{4}}(\tilde{n}\lambda_{1} - \tilde{n}\lambda_{2}) \nonumber \\
    & + \frac{2 p m^{2}_{\mu s}}{M^{4}}\tilde{n}\lambda_{3}, \nonumber \\
    (H^{\prime})^{(2)}_{\mu \tau} =& 0, \nonumber \\
    (H^{\prime})^{(2)}_{\tau \tau} =& \frac{m^{2}}{M^{2}} + \frac{p m^{2}}{M^{4}}(1 - \cos (2\phi_{12}))\tilde{n}\lambda_{1} + \frac{p m^{2}}{M^{4}}(1 + \cos (2\phi_{12}))\tilde{n}\lambda_{2} - \frac{2 p m^{2}}{M^{4}}\tilde{n}\lambda_{3}
    \label{eq:second_order_Hprime_phi12_2p1}
.\end{align}
As before, we use standard perturbation techniques to calculate the approximate energy eigenvalues. The leading order energy eigenvalues are the eigenvalues of $(H^{\prime})^{(0)}$ and is given in Eq.~\eqref{eq:leading_order_eigenval_phi12_2p1}. There are no first-order corrections to the energy eigenvalues. The second order correction are given in Eq.~\eqref{eq:second_order_eigenval_phi12_2p1}.

\section{\texorpdfstring{Calculation of the Hamiltonian for Sec.~\ref{subsec:phi13_2p1}}{Calculation of the Hamiltonian for Sec. phi13-2p1}}
\label{app:2p1_phi13}

In this appendix, we will give details of the calculation in Sec.~\ref{subsec:phi13_2p1}. Computing the Hamiltonian in the RHS of Eq.~\eqref{eq:schro_eqn_2p1} for $\phi_{13} \neq 0, \phi_{12} = \phi_{23} = 0$ and writing
\begin{align}
    H = p + \begin{pmatrix}H_{e e} & H_{e \mu} & H_{e \tau}\\ H^{*}_{e \mu} & H_{\mu \mu} & H_{\mu \tau}\\ H^{*}_{e \tau} & H^{*}_{\mu \tau} & H_{\tau \tau}\end{pmatrix}
,\end{align}
we get 
\begin{align}
    H_{e e} =& \frac{m^{2}_{e s}}{2 p} + V_{CC} + V_{NC} + \biggl[ \frac{\sin^{2}\phi_{13}m^{2}_{e s}}{M^{2}} + \frac{\sin (\alpha - \phi)\sin (2\phi_{13}) m_{e s}m_{\mu s}}{m M} + \frac{\cos^{2}\phi_{13}m^{2}_{\mu s}}{m^{2}} \biggr] \tilde{n}\lambda_{1} + \frac{m^{2}_{e s}}{m^{2}}\left( 1 - \frac{m^{2}}{M^{2}} \right)\tilde{n}\lambda_{2} \nonumber \\
    & + \biggl[ \frac{\cos^{2}\phi_{13} m^{2}_{e s}}{M^{2}} - \frac{\sin(\alpha - \phi)\sin(2\phi_{13})m_{e s} m_{\mu s}}{m M} + \frac{\sin^{2}\phi_{13}m^{2}_{\mu s}}{m^{2}} \biggr]\tilde{n}\lambda_{3}, \nonumber \\
    H_{e \mu} =& \frac{1}{2 m^{2} M^{2} p}  e^{i(\phi_{1}-\phi_{2})} \biggl(-2 p m_{e s}^{3} m_{\mu s} \tilde{n}\lambda_{2} - i m M p e^{-i(\alpha - \phi)} \sin(2 \phi_{13}) m_{e s}^{2} (\tilde{n}\lambda_{1}-\tilde{n}\lambda_{3}) \nonumber \\
    & -i e^{i(\alpha - \phi)} m M p \sin(2 \phi_{13}) m_{\mu s}^{2} (\tilde{n}\lambda_{1}-\tilde{n}\lambda_{3}) + m_{e s} m_{\mu s} \biggl(m^{2} M^{2} - p \bigl(-m^{2} + M^{2} + (m^{2}+M^{2}) \cos(2 \phi_{13})\bigr) \tilde{n}\lambda_{1} \nonumber \\
    & + p (M^{2}-m_{\mu s}^{2}) \tilde{n}\lambda_{2} + m^{2} p \tilde{n}\lambda_{3}\cos^{2}\phi_{13} - M^{2} p \tilde{n}\lambda_{3}\sin^{2}\phi_{13}\biggr)\biggr), \nonumber \\
    H_{e \tau} =& e^{i\phi_{1}}\biggl[ \frac{M m_{e s}}{2 p} + \frac{\sin^{2}\phi_{13} m_{e s}}{M}\tilde{n}\lambda_{1} - \frac{m_{e s}}{M}\tilde{n}\lambda_{2} + \frac{\cos^{2}\phi_{13}m_{e s}}{M}\tilde{n}\lambda_{3} - i e^{i(\alpha - \phi)}\sin (2 \phi_{13})\frac{m_{\mu s}}{2 m} (\tilde{n}\lambda_{1} - \tilde{n}\lambda_{3})  \biggr], \nonumber \\
    H_{\mu \mu} =& \frac{m^{2}_{\mu s}}{2 p} + V_{NC} + \biggl[ \frac{\sin^{2}\phi_{13} m^{2}_{\mu s}}{M^{2}} - \frac{\sin(\alpha - \phi)\sin(2 \phi_{13})m_{e s}m_{\mu s}}{m M} + \frac{\cos^{2}\phi_{13}m^{2}_{e s}}{m^{2}} \biggr]\tilde{n}\lambda_{1} + \frac{m^{2}_{\mu s}}{m^{2}}\left( 1 - \frac{m^{2}}{M^{2}} \right)\tilde{n}\lambda_{2} \nonumber \\
    & + \left( \frac{\cos^{2}\phi_{13} m^{2}_{\mu s}}{M^{2}} + \frac{\sin(\alpha - \phi)\sin(2\phi_{13})m_{e s}m_{\mu s}}{m M} + \frac{\sin^{2}\phi_{13}m^{2}_{e s}}{m^{2}} \right)\tilde{n}\lambda_{3}, \nonumber \\
    H_{\mu \tau} =& e^{i\phi_{2}} \biggl[ \frac{M m_{\mu s}}{2 p} + \frac{\sin^{2}\phi_{13}m_{\mu s}}{M}\tilde{n}\lambda_{1} - \frac{m_{\mu s}}{M}\tilde{n}\lambda_{2} + \frac{\cos^{2}\phi_{13}m_{\mu s}}{M}\tilde{n}\lambda_{3} + i e^{i(\alpha - \phi)}\frac{\sin (2\phi_{13})m_{e s}}{2 m}(\tilde{n}\lambda_{1} - \tilde{n}\lambda_{3}) \biggr], \nonumber \\
    H_{\tau \tau} =& \frac{M^{2}}{2 p} + \frac{m^{2}}{2 p} + \sin^{2}\phi_{13}\left( 1 - \frac{m^{2}}{M^{2}} \right)\tilde{n}\lambda_{1} + \frac{m^{2}}{M^{2}}\tilde{n}\lambda_{2} + \cos^{2}\phi_{13}\left( 1 - \frac{m^{2}}{M^{2}} \right)\tilde{n}\lambda_{3} \label{eq:H_phi13_2p1}
.\end{align}
We can again define an $H^{\prime} = (2 p H)/M^{2}$ and ignore the leading $p$. Collecting all the terms $\mathcal{O}((m/M)^{0})$, we get
\begin{align}
    (H^{\prime})^{(0)}_{e e} =& \frac{2 p}{M^{2}}(V_{CC} + V_{NC}) + \frac{2 p \cos^{2}\phi_{13}m^{2}_{\mu s}}{m^{2}M^{2}}\tilde{n}\lambda_{1} + \frac{2 p m^{2}_{e s}}{m^{2}M^{2}}\tilde{n}\lambda_{2} + \frac{2 p \sin^{2}\phi_{13}m^{2}_{\mu s}}{m^{2}M^{2}}\tilde{n}\lambda_{3}, \nonumber \\
    (H^{\prime})^{(0)}_{e \mu} =& - e^{i(\phi_{1}-\phi_{2})}\frac{2 p m_{es}m_{\mu s}}{M^{2}}(\tilde{n}\lambda_{2}-\tilde{n}\lambda_{3}), \nonumber \\
    (H^{\prime})^{(0)}_{e\tau} =& -ie^{i(\alpha -\phi_{2})}\frac{p \sin (2\phi_{13}) m_{\mu s}}{m M^{2}}(\tilde{n}\lambda_{1} - \tilde{n}\lambda_{3}), \nonumber \\
    (H^{\prime})^{(0)}_{\mu \mu} =& \frac{2 p}{M^{2}}V_{NC} + \frac{2 p \cos^{2}\phi_{13}m^{2}_{e s}}{m^{2}M^{2}}\tilde{n}\lambda_{1} + \frac{2 p m^{2}_{\mu s}}{m^{2}M^{2}}\tilde{n}\lambda_{2} + \frac{2 p \sin^{2}\phi_{13}m^{2}_{e s}}{m^{2}M^{2}}\tilde{n}\lambda_{3}, \nonumber \\
    (H^{\prime})^{(0)}_{\mu \tau} =& i e^{i(\alpha - \phi_{1})} \frac{p \sin(2\phi_{13})m_{e s}}{m M^{2}}(\tilde{n}\lambda_{1} - \tilde{n}\lambda_{3}), \nonumber \\
    (H^{\prime})^{(0)}_{\tau \tau} =& 1 + \frac{2 p \sin^{2}\phi_{13}}{M^{2}}\tilde{n}\lambda_{1} + \frac{2 p \cos^{2}\phi_{13}}{M^{2}}\tilde{n}\lambda_{3}
    \label{eq:leading_order_Hprime_phi13_2p1}
.\end{align}
The characteristic equation of $(H^{\prime})^{(0)}$ is a cubic equation and is given in Eq.~\eqref{eq:phi13_char_eqn}. The expression of energy eigenvalues is already quite complicated in the leading order. Hence we have not calculated the energy eigenvalues including higher-order corrections and the corresponding elements of the dimensionless Hamiltonian.

\bibliography{references}
\bibliographystyle{apsrev4-2}

\end{document}